\documentclass[iop]{emulateapj}
\usepackage{amsmath}
\usepackage{aas_macros}
\usepackage{enumitem}[0]

\usepackage[utf8]{inputenc}
\usepackage{tikz}
\usetikzlibrary{shapes.geometric, arrows}
\usepackage{verbatim}
\usepackage[normalem]{ulem}
\usepackage{tabu}

\newcommand{\yr}{\mbox{yr}}
\newcommand{\lyr}{\mbox{lyr}}
\newcommand{\Myr}{\mbox{Myr}}
\newcommand{\Gyr}{\mbox{Gyr}}
\newcommand{\kms}{\mbox{km/s}}
\newcommand{\pc}{\mbox{pc}}
\newcommand{\settleable}{settleable}
\newcommand{\eq}{\mbox{\tiny{eq}}}
\newcommand{\AU}{\mbox{AU}}

\shorttitle{The Fermi Paradox and the Aurora Effect}
\shortauthors{Carroll-Nellenback {\it et al.}}

\begin{document}

\title {The Fermi Paradox and the Aurora Effect: Exo-civilization Settlement, Expansion and Steady States}
\author{Jonathan Carroll-Nellenback}
\affiliation{Department of Physics and Astronomy, University of Rochester, Rochester, New York, 14620}
\author{Adam Frank}
\affiliation{Department of Physics and Astronomy, University of Rochester, Rochester, New York, 14620}
\author{Jason Wright}
\affiliation{Department of Astronomy and Astrophysics, The Pennsylvania State University, University Park, PA 16802}
\affiliation{Center for Exoplanets and Habitable Worlds, The Pennsylvania State University,  University Park, PA 16802}
\affiliation{PI, NASA Nexus for Exoplanetary Systems Science}
\author{Caleb Scharf}
\affiliation{Department of Astronomy, Columbia University, 550 West 120th St. New York, NY 10027, USA}
\affiliation{PI, NASA Nexus for Exoplanetary Systems Science}

\begin{abstract}
We model the settlement of the galaxy by space-faring civilizations in order to address issues related to the Fermi Paradox. We are motivated to explore the problem  in a way that avoids assumptions about the `agency' (i.e. questions of intent and motivation) of any exo-civilization seeking to settle other planetary systems. We begin by considering the speed of an advancing settlement front to determine if the galaxy can become inhabited with space-faring civilizations on timescales shorter than its age. Our models for the front speed include the directed settlement of nearby {\settleable} systems through the launching of probes with a finite velocity and range. We also include the effect of stellar motions on the long term behavior of the settlement front which adds a diffusive component to its advance. As part of our model we also consider that only a fraction $f$ of planets will have conditions amenable to settlement by the space-faring civilization.  The results of these models demonstrate that the Milky Way can be readily `filled-in' with settled stellar systems under conservative assumptions about interstellar spacecraft velocities and launch rates. We then move on to consider the question of the galactic steady-state achieved in terms of the fraction $X$ of settled planets.  We do this by considering the effect of finite settlement civilization lifetimes on the steady states.  We find a range of parameters for which $0<X<1$ i.e. the galaxy supports a population of interstellar space-faring civilizations even though some {\settleable} systems are uninhabited. In addition we find that statistical fluctuations can produce local over-abundances of {\settleable} worlds. These generate long-lived clusters of settled systems immersed in large regions which remain unsettled. Both results point to ways in which Earth might remain unvisited in the midst of an inhabited galaxy.  Finally we consider how our results can be combined with the finite horizon for evidence of previous settlements in Earth's geologic record.  Using our steady-state model we constrain the probabilities for an Earth visit by a settling civilization before a given time horizon.  These results break the link between Hart's famous ``Fact A" (no interstellar visitors on Earth now) and the conclusion that humans must, therefore, be the only technological civilization in the galaxy. Explicitly, our solutions admit situations where our current circumstances are consistent with an otherwise settled, steady-state galaxy. 

\end{abstract}


\section{Introduction}

\label{Introduction}
The Fermi Paradox has a long history in discussions of the prevalence of ``alien" technological civilizations (i.e. `exo-civilizations') in the galaxy \citep{Webb2002,Cirkovic2018}. Originating with a lunchtime conversation in 1950 where Enrico Fermi famously asked `where is everybody?' \citep{Jones1985}, the Fermi paradox was first formalized by  \citet{Hart1975} and has since become a standard framework for addressing questions concerning the prevalence of exo-civilizations \citep[but see][]{Gray2015}. Formally the paradox might be expressed as follows: ``If technologically advanced exo-civilizations are common, then we should already have evidence of their existence either through direct or indirect means" \citep{Frank2018}. Here we take {\it indirect} detection to mean the search for technosignatures \citep{Tarter07} from distant sources outside the solar systems \citep[e.g.][]{Siemion2013,Wright2014} while {\it direct} detection means material evidence for an exo-civilization's visit to Earth or our solar system \citep{Davies2012,Haqq-Misra2012,Wright2018b} 

Such a distinction between direct and indirect detection is important.  In Hart's formulation of the Fermi Paradox his ``Fact A" was the lack of aliens on Earth now. It was Fact A that then led Hart to conclude that no other technological civilizations any kind exist or have existed. The lack of indirect detection of exo-civilizations via radio or other signals represents a different constraint on alien life (But see \citet{Kuiper1977}, who use the idea that there should be probes in the Solar System as a reason to search for radio communication to those probes from abroad.) This apparent absence of signals has been been called a ``Great'' \citep{Brin1983} or ``Eerie" \citep{Davies2011} Silence. Such silence has been taken by some to serve as as its own answer to Fermi's Paradox (i.e.\ we don't see them because they don't exist). The assumption in this interpretation of the Paradox is that the Search for Extraterrestrial Intelligence \citep[SETI, e.g.][]{Tarter2001} has been extensive enough to place firm limits on the prevalence of exo-civilizations. This conclusion is, however, unwarranted.  \citet{Tarter2010} examined the volume of radio SETI search space, and concluded that only a tiny fraction of the radio SETI parameter space necessary to reach such conclusions has been covered. \cite{Wright2018a} amplified this conclusion with a similar calculation and concluded that the situation is equivalent to searching unsuccessfully for dolphins in a small pool's worth of ocean water and then concluding the ocean was dolphin-free.  

The Fact A interpretation of the Fermi Paradox, focusing on their presence on Earth (or at least in the solar system), presents greater difficulties in resolution. One of the first rigorous discussions of the possibility of contact via interstellar probes was that of \citet{Bracewell1960}, and \citet{Freitas1980} and \citet{Tipler1980} extended the idea to include self-replicating probes that would saturate the Galaxy. \citet{Hart1975} had in mind interstellar settlement by the intelligent species itself. Either way, the math is the same: both \citeauthor{Tipler1980} and \citeauthor{Hart1975} showed that sub-relativistic probes sent out by a single interstellar faring species would cross the galaxy in approximately $6.5\times 10^{5}$ y. Given that this is a small fraction of the galaxy's lifetime, it would seem that a single species intent on visiting or even settling the galaxy has had ample time to do so.  But \citet{Freitas1985} challenged this version of the Fermi Paradox as well, noting that it is not a formal paradox at all, both since it relies on the assumptions that alien life {\it would certainly} launch such probes, that those probes would be in the solar system today, {\it and} that we would have noticed them by now.  

A number of authors have attempted to explore these nuances. \citet{Ashworth2014} included settlement parameters such as maximum probe range and maximum travel speed among his categorization of ``solutions'' to the Fermi Paradox. Others have attempted to invalidate or verify the order-of-magnitude results of \citeauthor{Hart1975} and \citeauthor{Tipler1980} using both general arguments and more detailed models.

\citet{Stull1979} argued that competition among settled systems for the small number of frontier systems would stall the expansion front. \citet{Jones1976} argued that  a species might intentionally slow its expansion in the Galaxy by at least an order of magnitude via population restrictions.  \cite{Jones1978} expanded on this theme with a Monte Carlo approach, performing the calculation of a settlement wave numerically, under the assumption that the spread of life through the Galaxy is driven by population growth, and that only planets or stellar systems with large populations would spread to other stars. He found rapid progress of the interstellar settlement front with the front moving at 6\% of the speed of the ships themselves. Based on these results and the lack of material evidence of exo-civilizations on Earth, Jones conjectured that no interstellar civilizations had yet arisen, consistent with \citeauthor{Hart1975} and \citeauthor{Tipler1980}. Later calculations by \citet{Jones1981} exploring a wider range of population growth assumptions came to similar conclusions.  

On the other hand, \citet{Sagan1983} argued that the sorts of self-replicating probes imagined by \citeauthor{Tipler1980} would be inherently dangerous and uncontrollable, and therefore would not be constructed in the first place. \citet{ChybaHand} argued that self-replicating probes would be subject to evolution, mutation and predation much like life is, greatly complicating the analysis, a proposition explored numerically by \citet{Wiley2011}.

Regarding the \citeauthor{Hart1975} scenario, \citet{Newman1981} describe an analytic calculation that reproduced the results of \citet{Jones1978}, but found that under reasonable assumptions about low population growth rates, the progression of the wave could be slow and the time to populate the Milky Way would approach or even exceed its age. Roughly speaking, their argument is that to be long-lived, a civilization must have low population growth, but if they have low population growth, they will not settle nearby systems \citep[see also the ``sustainability solution'' of][]{Haqq-Misra2009}.

\citeauthor{Sagan1983} also suggested that the ``colonization phase'' of the civilization would necessarily be finite in duration, and found that for durations less than $3\times10^6$ y we should not expect Earth to have been colonized. They argued that since longer durations were not ``plausible,'' Fact A posed no significant challenge to the hypothesis that the Milky Way is endemic with space-faring life.

Tipler responded in a series of papers, prompting \citet{Sagan1983} to present a detailed rebuttal to Tipler and defense of the \citeauthor{Newman1981} calculation, referring to Tipler's position as the ``solipsistic approach to extraterrestrial intelligence.'' Here they defended their choices of parameters for population growth, including their assertion that only well-populated planets would launch new settlement ships, and that civilizations would have finite colonization lifetimes.

\citet{Walters1980} pointed out that the possibility of life spreading among the stars has implications for the Drake Equation \citep{Drake1965}, and computed that for a maximum travel time of $10^3$ years one should multiply $N$ by a factor $C\lesssim10$ to account for this, effectively arguing that the difficulty of interstellar travel would limit each civilization to no more than 10 additional systems.

Additional work has been performed by \citet{Landis1998}, who included maximum probe range in his percolation model and had some settlements permanently cease sending out probes. \citet{Kinouchi2001} used an analogy to the large number of uninhabited portions of the Earth to derive a simple model for the ``persistence'' of uninhabited regions of the Galaxy, a model with \citet{Galera2018} refined and expanded.

\citet{Bjork2007} and \citet{Hair2013} performed numerical calculations in a 3D grid of stars representative of the Galaxy. \cite{Cotta2009} used a 2D grid of stars to explore a two-stage colonization strategy with fast exploration probes and slower colonization probes. \citet{Gros05} explored the possibility that the settlement wave would cease for cultural reasons.  \citet{Zackrisson2015} explored the settlement patterns of in-progress galactic settlement to guide observational searches for Type III Kardashev civilizations.

\citet{Lin2015} modeled the spread of non-technological life via panspermia, concluding that spatial correlations among life-bearing exoplanets would provide strong evidence for the hypothesis.  \citet{Forgan2009} describes a general numerical model for the rise and spread of life in the Galaxy, suitable for testing a wide variety of Fermi Paradox-related hypothesis (see references therein). \citet{Vukotic2012} expand on \citeauthor{Forgan2009}'s work with cellular automata theory.

With the exception of \citet{Zackrisson2015}, who included both 3D stellar thermal motion and galactic shear in their calculations, all of these studies assumed that settlement occurs across a static substrate of stars, and most worked in 2D. As pointed out by \citet{Brin1983}, \citet{Ashworth2012}, and \citet{Wright2014}, this assumption is probably fine in the case of rapid settlement of the Galaxy by relativistic probes, but it cannot be used to support any conclusion that there are regions of parameter space in which settlement stalls or large uninhabited regions persist for long times in an otherwise inhabited Galaxy, because it assumes that settlements cannot be spread through the Galaxy by the motions of the stars themselves. This is particularly important in scenarios with short maximum probe lifetimes (meaning probes that are either slow or have short range).

\citet{Ashworth2012} and \citet{Wright2014} also repeated the admonition of \citet{Hart1975} against reaching for ``solutions'' to the Fermi Paradox that invoke a permanent lack of interest in settling nearby stars, as done by \citet{Newman1981} and \citet{Sagan1983}, which is an example of what \citet{Wright2014} dubbed the ``monocultural fallacy''. Such solutions invoke the unknown and unknowable intentions or motivations of exo-civilizations, and so unless a species goes {\it extinct} we should not suppose that any propensity for colonization should go to zero permanently for all settlements. 

\citet{Wright2014} also address the assumption of \citet{Newman1981} and \citet{Sagan1983} that the drive to settle new systems would be driven entirely by population pressure, since interstellar migration can hardly be expected to reduce overcrowding \citep{vonHoerner1975} and such motivations for settlement probably do not even describe most migrations of humans across the Earth.

In addition to the question of the settlement front speed, one can also ask about the steady state properties of a galaxy which has already been swept across by civilization-bearing probes.  If we assume that civilizations have finite lifetimes, then an eventual balance should be achieved between the settlement of currently empty worlds and the death of civilizations on previously settled worlds. This question bears directly on Hart's Fact A.  If civilizations have a finite duration, then it is possible that Earth was settled some time in the distant past and all traces of that settlement have been erased by geological evolution. In \cite{SchmidtFrank2018} it was shown that evidence of previous industrial civilizations in Earth's deep past would likely be found only in isotopic and chemical stratographic anomalies and that the experiments needed to pinpoint non-natural origins for such signals had yet to be performed.  The question of solar system artifacts of previous civilizations (alien or otherwise) has also been addressed in \citet{Davies2012},\citet{Haqq-Misra2012}, and \citet{Wright2018b}.

In this paper we focus on the Fermi Paradox in the form of Hart's Fact A.  We first reexamine the issue of settlement front speeds using both analytic and simulation based methods to track the settlement front in a realistic gas of stars. 

We next take up the issue of equilibrium models for galactic habitation.  Novel aspects of our study include the explicit inclusion of thermal stellar motions coupled with the possibility that not all worlds are inherently {\it settleable}. Often there is the assumption that any planet can be "terraformed" to the specific needs of the settling civilization.  But the idea that the purpose of probes is to build habitable settlements and that all stellar systems are viable targets for such settlements goes to the agency of an exo-civilization; in our work we therefore relax this assumption. In addition, some stars may host indigenous forms life, which may preclude settlement for practical or ethical reasons (see, for instance, \citet{Kuiper1977} who suggested the biology of Earth might be incompatible with that of would-by settling species.)  This theme was explored in (spoiler alert) the novel {\it Aurora} by Kim Stanley Robinson (\cite{Robinson2015}) in which even though a world was formally habitable it was not what we would call {\settleable}.  Thus we include the possibility that ``good worlds are hard to find" - what we call the {\it Aurora Effect} - via a pre-settlement fraction $f$ in our calculations. Separating the observable space density of stars from the unknown density of {\settleable} represents an important aspect of our study. 

Finally we consider the effect of finite lifetimes for civilizations which arise on the settled worlds in our calculations.  This allows us to include the possibility that Earth was settled by an exo-civilization at some time in the deep past but no evidence remains of its existence \cite{SchmidtFrank2018}). By including the temporal horizon over which evidence of such a settlement would persist, we are able to constrain the galactic equilibrium fraction.

The plan of the paper is as follows. In section 2 we introduce and explore an analytic model for the speed of the settlement front including stellar motions and the fraction of settle-able systems $f$.  In section 3 we present an agent-based numerical model for the evolution of a settlement front.  In section 4 we discuss the results of the numerical model and the implications for the long term evolution of the settlement of the galaxy.  In section 5 we discuss an equilibrium model for the galaxy fraction of settled systems and discuss the resulting implications given evidence implied by geological Earth's record. And in section 6 we present results of the steady state numerical models.
   
\section{Dynamic Model}
 We first consider the speed of a settlement front driven by the spread of `intelligent' agents (i.e. agents following a set of algorithmic rules) constrained by the limited range of the spacecraft and the dynamical diffusion of the stellar substrate.  In this model we assume that expansion proceeds via short-range “probes” which travel to a nearby system and “settle” it. A “settled’ system in this model takes on identical properties to the Ur system, launching additional probes to unsettled nearby system. 
 
 For simplicity we fix the maximum speed of settlement probes $v_p$ (relative to their host systems) as well as the maximum distance a probe can travel $d_p$ in the rest frame of the host system.  We also model the stellar substrate as having a Maxwellian velocity distribution with an average velocity of $v_s$, and a mean density of systems $\rho$ of which some fraction $f$ are {\settleable}.  Probes can be launched from a system with a periodicity $T_p$. For each probe launched the intelligent agents target the uninhabited system with the shortest travel time.  We also assume for simplicity that systems once settled continue to be so, although later in this study we will consider steady state models in which settled systems have finite lifetimes.

We can scale the model in terms of the probe range $(d_p)$, velocity $(v_p)$, and travel time $(t_p \equiv d_p/v_p)$ to reduce the 5 parameters described above into 3 dimensionless quantities:

\begin{align}
\eta=&f \rho d_p^3 \\
\nu_s=&\frac{v_s}{v_p} \\
\tau_p=&\frac{T_p}{t_p} \\
\end{align}

$\eta$ is the normalized density of {settleable} systems and roughly corresponds to the expected number of systems within range at any given point. $\nu_s$ is the relative speed of stellar substrate motions to probe motion and tracks the importance of the velocity of stars in aiding or restricting galactic settlement.  $\tau_p$ is the ratio of probe launch period to the probe travel time and corresponds to the relative delay due to building probes before they can be launched.  For reference, \cite{Henry2018} estimate the density of stars in the solar neighborhood between $.07\pc^{-3}$ and $.09\pc^{-3}$.  Assuming $\rho = .08 \pc^{-3} = 0.0023 \lyr^{-3}$ and $v_s=30 \kms$, we have $\eta = 2.3 f \left (\frac{d_p}{10 \lyr}\right)^3$, $\nu_s=.01 \left (\frac{v_p}{0.01c} \right)^{-1}$, and $\tau_p=.1 \left (\frac{T_p}{100 \yr} \right)\left (\frac{v_p}{0.01c} \right)\left (\frac{d_p}{10 \lyr}\right)^{-1}$.  We then seek a solution to the  motion of the agents expansion front $r(t)=vt$, or in scaled units $\xi(\tau)=\nu \tau$ where $\tau = \frac{t}{t_p}$, $\xi = \frac{r}{d_p}$, and $\nu = \frac{v}{v_p}$.  A summary of the parameters of our model is provided in table \ref{tableofparams}.

\begin{table*}[t]
\centering
\begin{tabular}{ | c| c | l | }
\hline
\textbf{Symbol} & \textbf{Definition} & \textbf{Description} \\ \hline
    $f$  & {\bf (reserve this column 4 defs not results)} & Fraction of systems that are {\settleable} \\ \hline
    $\rho$ & & Density of systems\\ \hline
    $d_p$  & & Probe range \\ \hline
    $v_p$ & & Probe velocity \\ \hline
    $v_s$ & & Average velocity of stellar substrate \\\hline
    $T_p$ & & Probe launch period (equivilent to the probe assembly time) \\\hline
    $t_p$ & $d_p/v_p$ & Probe travel time \\\hline
    $T_c$ & $\left ( \pi f \rho d_p^2 v_s \right)^{-1}$ & Encounter time between systems due to stellar motions \\\hline
    $T_l$ & $\mathcal{D}_1 T_p + (1-\mathcal{D}_1) T_c$ & Effective probe launch period \\ \hline
    $T_s$ & & settlement civilization lifetime \\ \hline
    $\eta$ & $f \rho d_p^3$  & Normalized density of {settleable} systems within probe range \\\hline
    $\nu_s$ & $v_s/v_p$ & Velocity of stellar substrate normalized by probe speed\\\hline
    $\tau_p$ & $T_p/t_p$ & Probe launch period normalized to probe travel time \\\hline
    $\tau_c$ & $\left(\pi \eta \nu_s\right)^{-1}$ & Encounter time normalized to probe travel time\\\hline
    $\mathcal{D}_1$ & $1-e^{-\frac{4 \pi }{3} \epsilon \eta}$ & Odds of another unsettled system being in range and ahead of settlement front \\ \hline
    $\tau_l$ & $\mathcal{D}_1 \tau_p + (1-\mathcal{D}_1) \tau_c$ & Expected probe launch period normalized to probe travel time. \\ \hline
    $\epsilon$ & $\frac{1}{4}$ & Odds of an upwind system being unsettled (parameter)\\ \hline
    $\nu_l$ & $1+3\tau_l^3 \log \frac{\tau_l}{\tau_l+1} + 3 \tau_l^2 - \frac{3}{2}\tau_l$ & Average {\it effective} probe velocity normalized to probe velocity\\ \hline
    $\mathcal{V}$ & 2 - 3.5 & Ratio of fastest speed to average speed for Maxwell-Boltzmann system \\ \hline
    $\nu$ & $\max \left [ \mathcal{V}\nu_s,\nu_l \right ]$ & Front speed normalized to probe speed \\ \hline
    $\Delta \xi$ & $\nu \tau_l/\ln 2 $ & Front thickness normalized to probe range \\
\hline
\end{tabular}
\caption{Table of parameters}
\label{tableofparams}
\end{table*}

\subsection{Approximation of the expansion speed}
We next derive the expected front speed $\nu$ as a function of normalized density $\eta$, substrate velocity $\nu_s$, and launch period $\tau_p$ by first considering various limiting cases.
\subsection{High Stellar Density Limit}

We first consider the simplest case where the normalized density of {\settleable} systems $\eta >> 1$, so that there are always plenty of neighboring systems within range.  In the static limit for the stellar substrate, $\nu_s \rightarrow 0$, we would expect expansion to simply occur as fast as the probes can travel (including the time needed to built the next probe).  The time to travel the probe range $d_p$ and launch another probe would be $\frac{d_p}{v_p} + T_p$.  This gives a front speed $v = \frac{d_p}{d_p/v_p + T_p}$ or in our scaled units $\nu = \frac{1}{1+\tau_p}$.  This assumes that systems are evenly spaced at the probe distance.  If we assume that the probe destination is another system randomly located within the sphere of radius $d_p$, the average distance travelled by probes per trip in our scaled units $\xi = \frac{d}{d_p}$ will be the volume averaged radius
\begin{equation}
    \left< \xi \right> = \displaystyle \frac{3}{4 \pi} \int d\Omega \int_0^1 \xi^3 d\xi = \frac{3}{4}
\end{equation}

The average speed (including the probe launch period) in our scaled units $\nu_p = \frac{v}{v_p}$ will then be

\begin{align}
    \nu_p= & \left < \frac{\xi}{\xi+\tau_p} \right> = \displaystyle \frac{3}{4 \pi} \int d\Omega \int_0^{1} \frac{\xi}{\xi+\tau_p} \xi^2 d\xi \\
      = & 1+3\tau_p^3 \log \frac{\tau_p}{\tau_p+1} + 3 \tau_p^2 - \frac{3}{2}\tau_p \\
      \rightarrow & \frac{2/3}{2/3+\tau_p} \text{for } \tau_p << 1
\end{align}

Note the last line means that in the limit $\tau_p << 1$ the probe speed is the same as what would be expected for uniform trips of an effective distance $2/3 d_p$.  And when $\tau_p \rightarrow 0$, the front speed is just the probe speed and $\nu_p \rightarrow 1$. However, if probes are slow enough or the time to launch is long enough, the stellar substrate may actually move faster $\nu_s >> \nu_p$, in which case the stellar diffusion would control the rate of expansion. When this is the case the fastest stars, in the tail of the distribution, will determine the expansion rate.  Combining these possibilities we can estimate the front speed in the high stellar density limit as $\nu=\max \left [\nu_s, \nu_p \right]$

\subsection{Low Stellar Density Limit}

At low densities $\eta << 1$, host systems do not typically have other {\settleable} systems in range ($d_p$) and must wait for the background stellar substrate motions to bring them within range.  The frequency of these close encounters can be thought of as collision rates of particles with a radius of $\frac{d_p}{2}$.  For particles to `collide' the total distance between them must be twice their radius - or $d_p$.  This gives a collisional cross section $\sigma = \pi d_p^2$, and an average encounter time $T_c = \left(f \rho v_s \sigma \right)^{-1}$, or in dimensionless units $\tau_c\equiv \frac{T_c}{t_p}= \left (\pi \eta \nu_s \right)^{-1}$.  Defining a dimensionless {\it effective probe launch period} $\tau_l$ we can, therefore, not exceed this encounter period to get  $\tau_l = \max \left[\tau_c, \tau_p \right]$. Combining this with the high density limit gives a resulting front speed of 
\begin{align}
    \nu_l=  1+3\tau_l^3 \log \frac{\tau_l}{\tau_l+1} + 3 \tau_l^2 - \frac{3}{2}\tau_l
\end{align}

Also, note that when the density drops below $\eta < \frac{1}{\pi}$, the collision time becomes longer than the stellar drift time $\tau_c > \nu_s^{-1}$ or equivalently $\frac{d_p}{T_c} < v_s$.  In this limit, probes, regardless of their speed or assembly time, can only advance a distance $d_p$ per encounter time $T_c$.  In this limit, the maximum effective probe speed $\frac{d_p}{T_c}$ is less than the stellar drift speed $v_s$ - so probe launches/relaunches cannot outpace the stellar motions. Even a probe launched forward from the fastest moving system at the leading edge of the front - will land on a system drifting back towards the front - and it won't be able to find another system in range fast enough to further advance the front.

\subsection{Static Limit}
Before discussing intermediate system spatial densities with a dynamic stellar substrate, it is instructive to first consider the static limit $\nu_s=0$.  In this case the front propagation speed is limited by the effective probe speed (including launch times) $\nu=\nu_p$.  However if the density of {\settleable} systems $\eta$ drops below a critical density $\eta_c$, the expansion of the front can be halted due to insufficient connectivity among neighbors.

\subsubsection{Critical Density}
\begin{figure}

\includegraphics[width=\columnwidth]{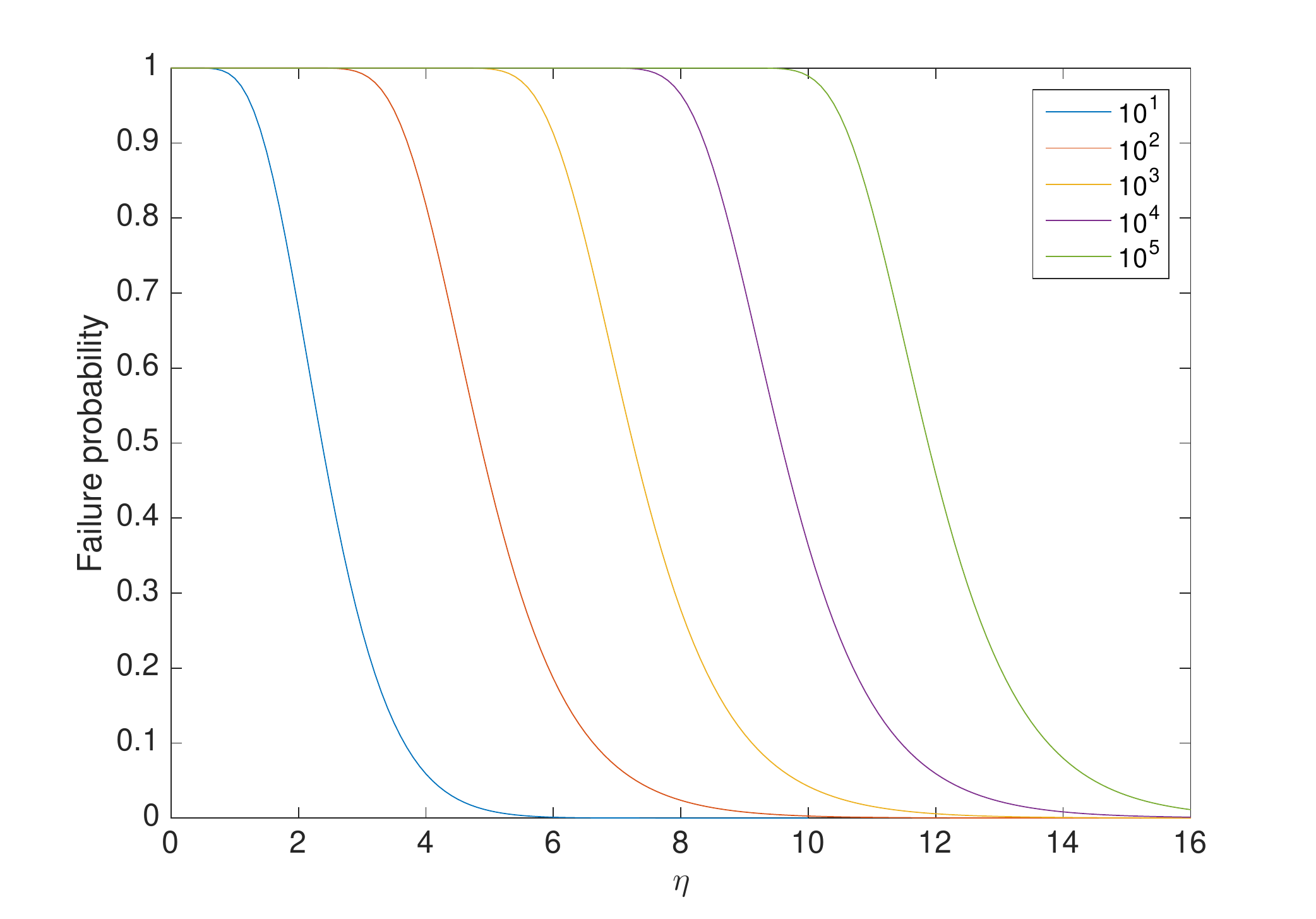}
\caption{Failure probability for a settlement front to completely cross a 1-D domain vs the average distance between stellar systems.  The average distance can be expressed as a mean "gap size", which in 1-D can be expressed in terms of the density of systems $\eta$.  As the distance that must be traveled between neighboring systems increases, the probability of the front advancing drops from $\sim 1$ to $\sim 0$. We also show how failure probability changes for domains with different numbers of systems $N$.  As the number $N$ of systems grows, larger average gap size can be tolerated in terms of keeping the settlement front from stalling because there are more neighbors that have gaps less than the probe range.
 \label{1dFailureProbability}}
\end{figure}

We first examine the effects of neighbor system connectivity via probes. In other words what are the dynamics of probes ``hopping" from neighbor to neighbor. We begin by considering the 1-dimensional equivalent of N systems distributed at random along the unit line (so that $\rho = N$) and ask what is the minimum probe range $d_p$ needed to ensure that no gaps exist that exceed the probe range, thereby halting the settlement front.  If $N >> 1$, the gaps are very weakly correlated and the gap sizes will have a beta distribution $\mathcal{B}(\alpha=1, \beta=N-1)$.   The odds that any gap is smaller than $d_p$ is given by the CDF (Cumulative Distribution Function) of the beta function which is the regularized incomplete beta function $I_{d_p}\left(\alpha=1,\beta=N-1 \right)$.  And the odds that no gap is larger than $d_p$ is 1 minus the odds that every gap is smaller than $d_p$ and is given by $1 - I_{d_p}\left(\alpha=1,\beta=N-1 \right)^N $.  Figure \ref{1dFailureProbability} shows what we call the failure probability $P_f$ for the settlement front as a function of the normalized density (1D equivalent) $\eta = \rho d_p = N d_p$.  This is defined as the odds of any gap size between systems being larger than than the probe range.  For normalized densities $\eta$ such that $P_t(\eta) \sim 1$ the settlement front can cross the domain. We also show the effect on $P_t$ of adding more systems to a domain of constant size.


In 3D, a single gap cannot stall the front, but in the static limit, we can treat the systems as a homogeneous Poisson point process with a volumetric rate parameter $\lambda = \rho$ which gives an occurrence rate for a volume of radius $d$ of $\Lambda = \frac{4}{3} \pi \rho d^3 = \frac{4}{3} \pi \eta \xi^3$ where we define a normalized distance $\xi = \frac{d}{d_p}$.  This gives the probability of finding $N$ neighbors within some normalized distance $\xi$
\begin{equation}
P_N(\xi)=\frac{\Lambda^N}{N!} e^{-\Lambda} = \frac{ \left ( \frac{4}{3} \pi \eta \xi^3 \right)^N}{N!} e^{-\frac{4}{3} \pi \eta \xi^3}
\end{equation}
The probability of having 1 or more neighbors within a distance $\xi$ is given by
\begin{equation}
    D_1(\xi) = 1 - P_0(\xi)
\end{equation}
And the probability of having $N$ or more neighbors within a normalized distance $\xi = \frac{d}{d_p}$ is given by
\begin{equation}
    \displaystyle D_N(\xi)=1 - \sum_{i=0}^{N-1} P_i(\xi)
\end{equation}
The differential change in the probability of finding $N$ or more neighbors as a function of $\xi$ is equal to the probability of finding the $N^{th}$ nearest neighbor at a distance $\xi$.  Taking the derivative we get the probability of finding the $N^{th}$ nearest neighbor.
\begin{equation}
  \displaystyle \mathcal{P}_N(\xi) = \frac{d D_N(\xi)}{d\xi} = - \sum_{i=0}^{N-1} \frac{d P_i(\xi)}{d\xi}
\end{equation}

We can use the recurrence relation

\begin{equation}
    \frac{d P_i(\xi)}{d\xi} = 
    \begin{cases}
    4 \pi \eta \xi^2 (P_{i-1}(\xi)-P_i(\xi)) & \text{if } i > 0 \\
    -4 \pi \eta \xi^2 P_i(\xi) & \text{if }i = 0 \\
    \end{cases}
\end{equation}
to write the probability of finding the $N^{th}$ nearest neighbor at a distance $\xi$
\begin{equation}
  \displaystyle \mathcal{P}_N(\xi) = -\sum_{i=0}^{N-1} \frac{d P_i(\xi)}{d\xi} =4 \pi \eta \xi^2 P_{N-1}(\xi)
\end{equation}
and then calculate the average distance to the $N^{th}$ nearest neighbor by taking the mean of the distribution
\begin{align}
  \displaystyle l_N = & \displaystyle \int_0^\infty \xi \mathcal{P}_N(\xi) d\xi = \displaystyle \int_0^\infty 4 \pi \eta \xi^3 P_{N-1}(\xi) d\xi \\
   = & \left ( \frac{4 \pi \eta}{3} \right)^{-1/3} \frac{\Gamma(N+\frac{1}{3})}{(N-1)!}
\end{align}
We can then determine what value of $\eta$ is required so that the average distance to the $N^{th}$ nearest neighbor is equal to the probe range $(l_N=1)$.
\begin{equation}
    \eta_N=\frac{3}{4 \pi} \left (\frac{\Gamma \left(N+\frac{1}{3} \right)}{\left(N-1\right)!} \right)^3 
\end{equation}
From this relation we find that when $\eta \ge \eta_1 \approx .1700$, the average distance to the nearest neighbor is within the probe range.  When $\eta \ge \eta_2 \approx .4030$ the average distance to the second nearest neighbor will be within the probe range, and when $\eta \ge \eta_3 \approx .6399$, the average distance to the third nearest neighbor will be within the probe range. 

These relations allow us to understand how the availability of multiple hops leads to a fully connected space of {\settleable} systems. To that end we consider how the typical number of connected systems changes as a function of the normalized density $\eta$.  To determine this we created a set of random points in a 3D box (with periodic boundary conditions) and then calculated the sizes of each isolated sub-region $N_i$ for various values of $\eta$.  In this the average number of accessible systems (not counting oneself) is given by 

\begin{equation}
N_a = \frac{\displaystyle \sum N_i (N_i-1) }{\sum{N_i}}
\end{equation}

\begin{figure}
\includegraphics[width=\columnwidth]{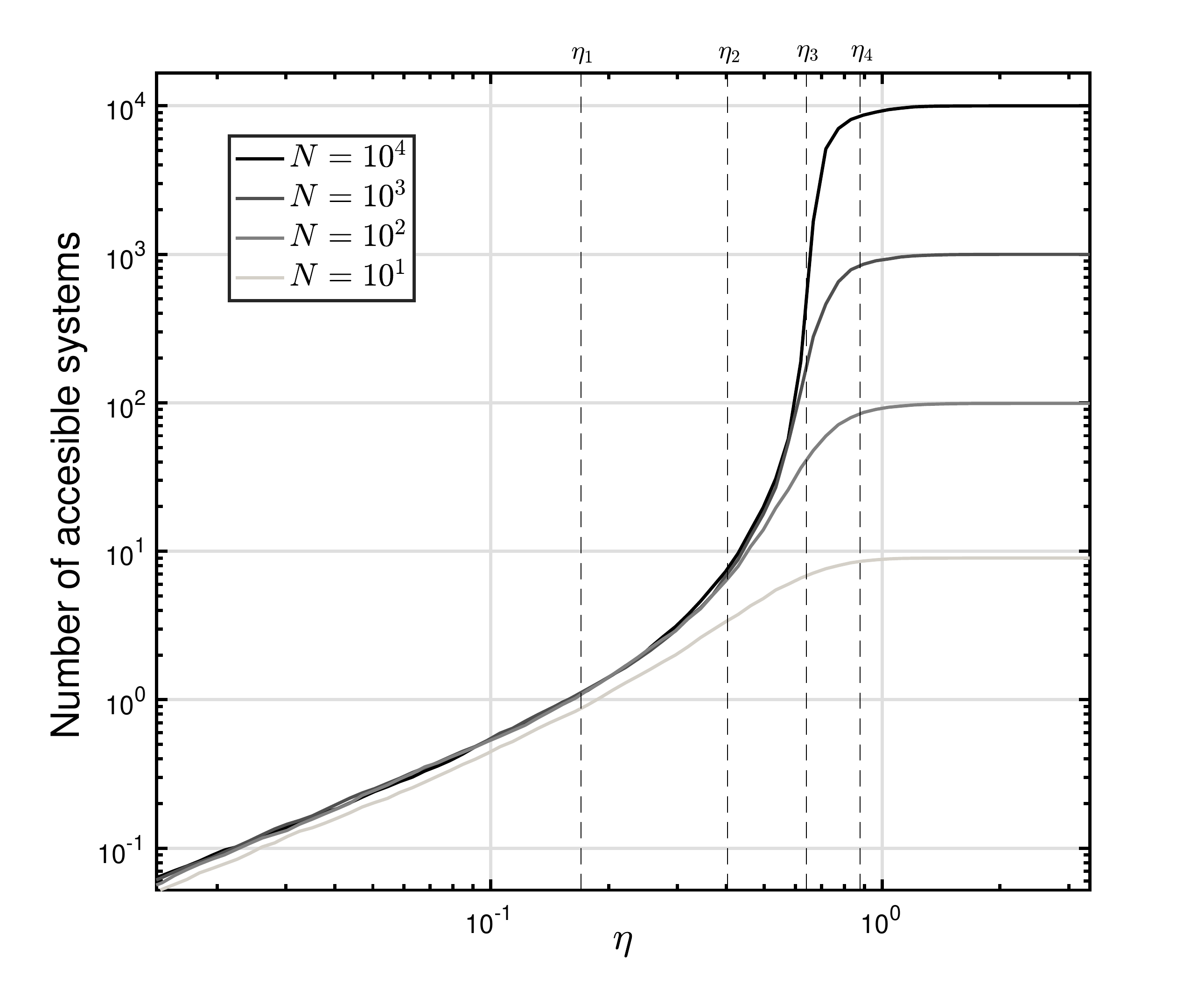}
\caption{The average number of accessible {\settleable} systems vs the space density of such systems. The density is normalized to the probe range $(\eta)$.  Note that systems tend to have a single {\settleable} neighbor when the average nearest neighbor distance equals the probe range $\eta = \eta_1$.  The space becomes fully connected, meaning a single original civilization can settle all {\settleable} systems, when the average distance to the $4^{th}$ nearest neighbor equals the probe range $\eta = \eta_4$. \label{3DClusterSize}}
\end{figure}

Figure \ref{3DClusterSize} shows the resulting average number of accessible systems as a function of density for $N=10^1$, $10^2$, $10^3$, \& $10^4$.  For $\eta \gtrsim \eta_1$, systems have on average a handful of other accessible systems. Once $\eta \gtrsim \eta_4 \approx .8777$, systems are nearly fully connected.  This means and the number of accessible systems for settlement is only limited by the total number of systems in the box.  Thus $\eta > \eta_c \equiv \eta_4 \sim .88$ represents a threshold density for settleable systems past which the settlement front should expand freely with the probe speed

\subsection{Intermediate Density}
Finally we consider the transition from the low density limit $\eta < \frac{1}{\pi}$ (stellar diffusion limited) to the high density limit $\eta > \eta_4$ (probe speed limited).  For intermediate densities, the actual time for probe launches will either be $\tau_p$, if at least one other unsettled system is in range, or it collision time $\tau_c$.   The odds that there is another unsettled system in range and positioned to advance the front
\begin{equation}
    \mathcal{D}_1=1-e^{-\frac{4 \pi }{3} \epsilon \eta}
\end{equation}
where $\epsilon$ is the product of the fraction of the volume that will assist in advancing the front and the fraction of systems in that volume that are not already settled.  For the system to be upwind of the front, $\epsilon < 1/2$.  And for our models, we find reasonable agreement with $\epsilon=\frac{1}{4}$. 

Using this we can refine our effective launch time that would advance the settlement front from $\tau_l=\max\left[\tau_p, \tau_c \right]$ to 
\begin{equation}
\tau_l =  \mathcal{D}_1 \tau_p + (1-\mathcal{D}_1) \tau_c
\end{equation}

\subsection{Stellar Velocity Distributions}
One final caveat, is that while $\nu_s$ is the average stellar speed, the front in the low density limit will be driven by the fastest stars.  Given $N$ stars with velocities taken from the Maxwell-Boltzmann distribution, the fastest moving star will be travelling a few times $v_s$ at $\mathcal{V}(N) v_s$ where $\mathcal{V}(N)$ is the expectation value for the maximum of $N$ systems taken from a Maxwell-Boltzmann distribution.  This factor scales fairly weakly with $N$ going from $2$ to $3.5$ for values of $N$ from $10^2$ to $10^5$.
This then gives our final model for the front speed
\begin{eqnarray}
\nu=\max \left [\mathcal{V} \nu_s, \nu_l \right ] \\
\nu_l=  1+3\tau_l^3 \log \frac{\tau_l}{\tau_l+1} + 3 \tau_l^2 - \frac{3}{2}\tau_l \\
\tau_l=\mathcal{D}_1 \tau_p + (1-\mathcal{D}_1) \tau_c \\
\tau_c=\frac{1}{\pi \eta \nu_s} \\
\mathcal{D}_1=1-e^{-\frac{4 \pi }{3} \epsilon \eta}
\end{eqnarray}

\subsection{Front Thickness}
After the leading edge of the front passes a given point, the local fraction of settled systems to total systems $X$ in the average rest frame should grow exponentially until it saturates. The growth rate should be proportional to the frequency of encounters between settled and unsettled systems.  The average  encounter frequency (normalized to the probe travel time) is just $\frac{1}{\tau_l}$ and the average encounter frequency between one settled and one unsettled system is $\frac{1}{\tau_l} X \left(1-X \right)$.  This gives the doubling rate - which corresponds to a continuous growth rate of the fraction of settled systems of $\frac{\ln 2}{\tau_l} X \left (1-X \right)$.  The time evolution of the settled fraction can therefore be written
\begin{equation}
    \frac{d X}{d\tau} =\frac{\ln 2}{\tau_l} X\left(1-X\right)
\end{equation}
This results in a logistic growth with a growth timescale of $\frac{\tau_l}{\ln 2}$.  In the frame of the front, this exponentially growth is stretched out spatially by the front speed $\nu$.  We expect the front to grow from unsettled to fully settled following a logistic curve with dimensionless width $\Delta \xi=\frac{\nu \tau_l}{\ln 2}$.  Note in the low density limit, we have $\tau_l = \left ( \pi \eta \nu_s \right)^{-1}$ and the front speed $\nu = \mathcal{V} \nu_s$.  This gives a width $\Delta \xi = \frac{\mathcal{V}}{\pi \ln 2} \eta^{-1} \approx \eta^{-1}$.

\section{Numerical Model}
To test our analytic model we ran a suite of numerical agent based simulations.  During each time-step, settled systems check to see if they are ready to launch a probe. Systems that are ready to launch a probe will target the unsettled system with the shortest intercept time subject to the constraint that the distance to intercept (in the systems frame of reference) is less than the probe range and that the time to intercept is less than the probe travel time.  If the time to intercept is longer than the probe travel time, a system waits to launch the probe until the time to intercept is less than or equal to the probe travel time.  Probes are not preemptively launched at sub probe speeds towards intercept locations - as those locations can be reached with a shorter trip duration by simply waiting to launch a probe at the probe speed provided the probe has not been launched toward another system in the meantime - or enough time has passed to have built another.  This also avoids systems preemptively targeting other systems that they would not be able to settle for a long time - allowing those systems to potentially be settled sooner.  Once a system is targeted, it will not be targeted by other probes, and will become settled after the probe intercept time.  

\begin{figure*}[t]
\centering
\includegraphics[width=\textwidth]{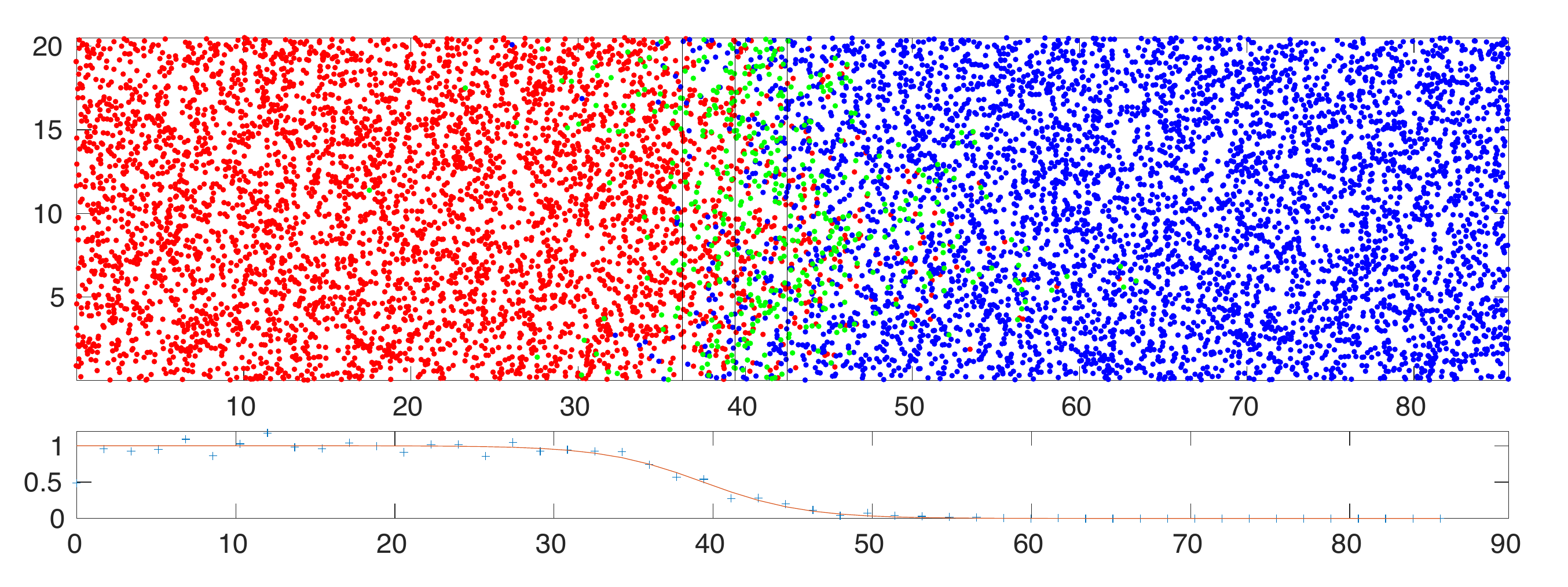}
\caption{Agent based model simulation of the expansion of a galactic settlement front after reaching steady state. The top panel shows a projected 2D snapshot of a 3D simulation with the density normalized to the probe range ($\eta=.2783$).  Red circles correspond to settled systems, green to targeted systems, and blue to unsettled systems. The settlement "front" is apparent in the transition from red to blue. The bottom panel shows the 1D fraction of settled systems along the direction of front propagation (+x) and the logistic fit $\left [ 1+\exp{\left (\frac{\xi'-\xi}{\Delta \xi}\right )} \right]^{-1}$ used to determine the position $(\xi)$ and width $(\Delta \xi)$ of the front.
\label{frontimage}}
\end{figure*}

\subsection{Initial Setup}
Initially the systems are randomly distributed within a periodic box with velocities taken from a Maxwell-Boltzmann distribution.  The initial distribution of settled systems is a Heaviside function $\mathcal{H}(x_0-x)$.  This creates a gradient in settled systems that causes the 'front' to naturally propagate to the right ($+x$).  To follow the front over many crossing times, the reference frame is also shifted into one moving to the right at the expected front velocity.  Settled systems that leave the left boundary are 'reused' and become unsettled as they wrap around and re-enter through the right boundary.  Initially the speed of this front is estimated using the analytic prescription, but if the front comes within the probe range of either the left or right boundary, the simulation is stopped, the front speed is re-estimated using the result of the simulation, and the process repeats.  

One final caveat is that in the low density diffusion limited regime, the system with the fastest velocity in the +x direction - will eventually reach the front. As it crosses the front it will settle additional systems that it encounters leaving a cone shaped wake of settled systems and locally increasing the speed of the front.  In a truly infinite plane parallel model - the front would have several 'fast' stars located at different locations along the front causing the front to be somewhat corrugated as the overlapping wakes from the fastest stars intersect.  In the simulations, the transverse periodic boundaries effectively limit the "corrugation scale" - and the normal periodic boundary conditions ensure that the fastest moving star eventually emerges from the front - at which point the front speed naturally increases to match the fastest particle's speed.  In our setup, we therefore shift the systems so that the system with the fastest velocity in the +x direction starts at $x_0$ at the leading edge of the front.  This position is chosen to be 2/3 of the distance across the box.  

The simulation box volume was chosen to contain $N=10^4$ habitable systems and simulations were performed for various values of $\eta = [10^{-1} , 10^1]$, $\tau_p=[0.1, 1.0]$, and $\nu_s=[10^{-3},10^{-1}]$.  The volume (in units of $d_p^3$) is given by $\frac{N}{\eta}$ which using our standard probe range $d_p=10 \lyr$ corresponds to volumes ranging from $\left (100 \lyr \right)^3$ to $\left ( 464 \lyr \right)^3$ and densities of habitable systems ranging from $.04$ to $4$ times the density of stars in the solar neighborhood.  The extents in $x$, $y$, and $z$ (in units of $d_p$) were $[w+15\Delta \xi, w, w]$ where $\Delta \xi = \nu \tau_l$ was the approximated front width which varied from $4.6$ to $117 \lyr$ and $w$ was solved for using the volume constraint.   For the various parameters, $w$ varied from $80 \lyr$ to $224 \lyr$ and in all cases $w$ was at least 8 probe travel distances as well as at least $10$ times the system separation scale $\eta^{-1/3}$ which went from $4.6 \lyr$ to $22 \lyr$.  In addition if we assume background stellar motions of $30 \kms$, the probe speed varied from $10^{-3}c$ to $10^{-1} c$ giving probe travel times that varied from $100$ to $10000 \yr$ and probe assembly times that varied from $1$ to $1000 \yr$. The simulations were each run for the longer of $100$ effective launch periods $(\tau_l)$ or $10$ probe travel times and varied from $1000 \yr$ to $30 \Myr$ and was sufficient for the system to have reached a steady state.  

\section{Results}
The top panel of figure \ref{frontimage} shows a typical snapshot of the numerical simulation (projected to 2D).  The red circles correspond to settled systems while the blue circles are unsettled.  Green systems are also unsettled, but have been targeted by a settled system.  The lower panel shows the fraction of settled systems projected onto 1d as well as the fit to the logistic curve $X=\left [ 1+\exp{\left (\frac{\xi'-\xi}{\Delta \xi}\right )} \right]^{-1}$ where $X$ is the fraction of systems that are settled, $\xi'$ is the dimensionless position (normalized to $d_p$ in the approximately co-moving frame), and $\xi$ and $\Delta \xi$ are fit parameters that indicate the dimensionless position and thickness of the front (in the co-moving frame).  The change in the dimensionless position $\xi$ in the approximately co-moving frame is then used to measure the front velocity.  For each run, we then calculate the location and average thickness from $20 \tau_l$ to $100 \tau_l$ and calculate the average front speed and thickness over this time period.

\begin{figure}
\centering
\includegraphics[width=1\columnwidth, bb=35 0 456 576, clip=true]{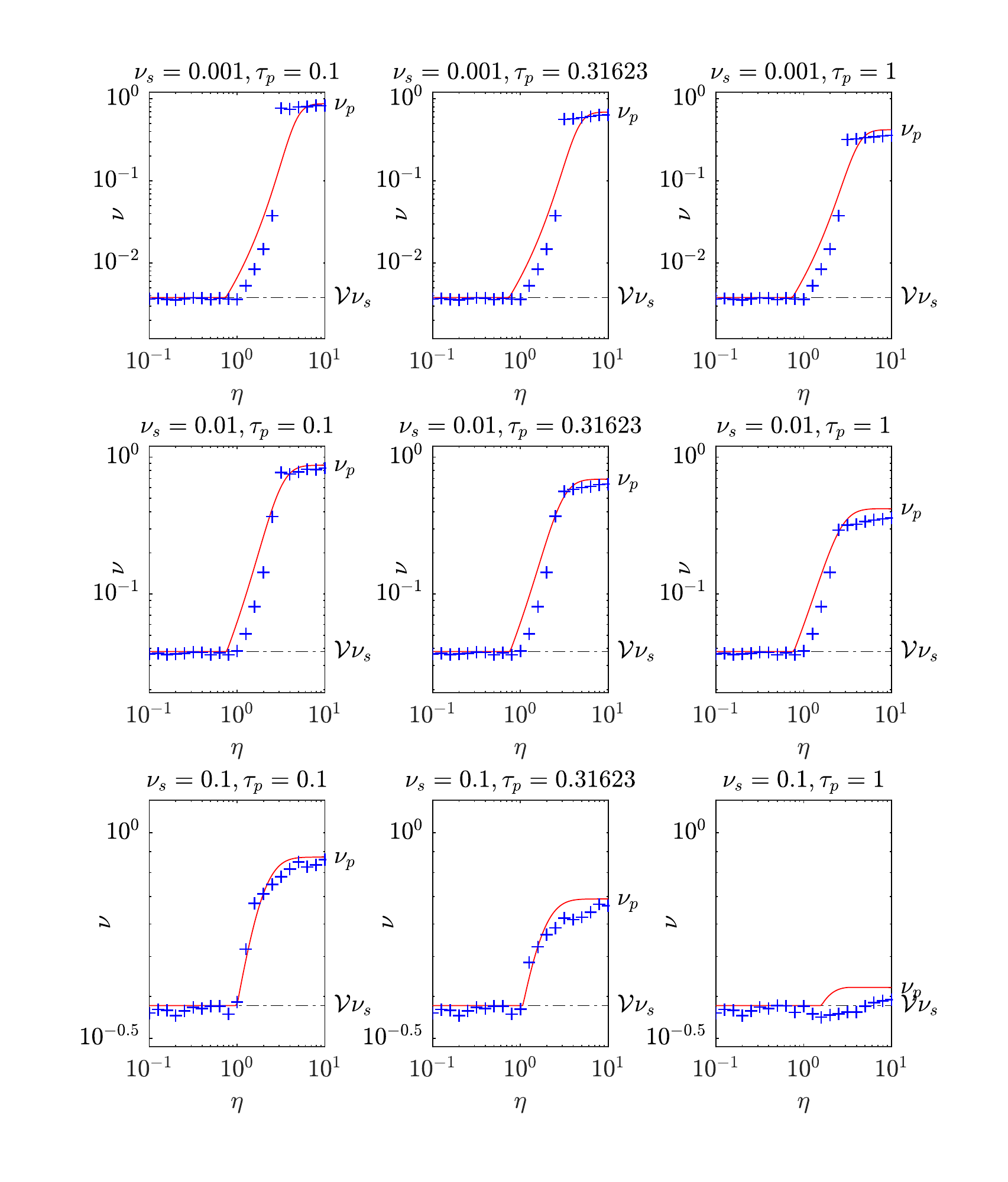}
\caption{Comparison of analytic and simulation results for the settlement front propagation speed. This figure shows 1D front speeds vs normalized density.  The red line comes from the analytic model and the blue dots come from the simulations. We show runs with: the ratio of stellar speeds to the probe speed $\nu_s=[.001,.01,.1]$; the ratios of probe relaunch times to probe travel time $\tau_p=[.1,.31623,1]$.  As $\eta$ increases, the front speed goes from diffusion limited, to collision limited, to probe speed limited. This figure (and the next) demonstrate the general accuracy of the analytic model discussed in the text.
\label{SpeedResults}}
\end{figure}

Figure \ref{SpeedResults} shows the resulting front speed from the numerical model over a range of values for $\eta$ for various combinations of $\nu_s$ and $\tau_p$ as well as the analytic estimate. In the low density region, the front speed is just given by $\mathcal{V}\nu_s$.  In the intermediate density region, there is a transition from $\mathcal{V}\nu_s$ to $\nu_p$ as the effective launch period $\tau_l$ transitions from the encounter time $\tau_c$ to the probe assembly period $\tau_p$.   

\begin{figure}
\centering
\includegraphics[width=1\columnwidth, bb=35 0 458 579, clip=true]{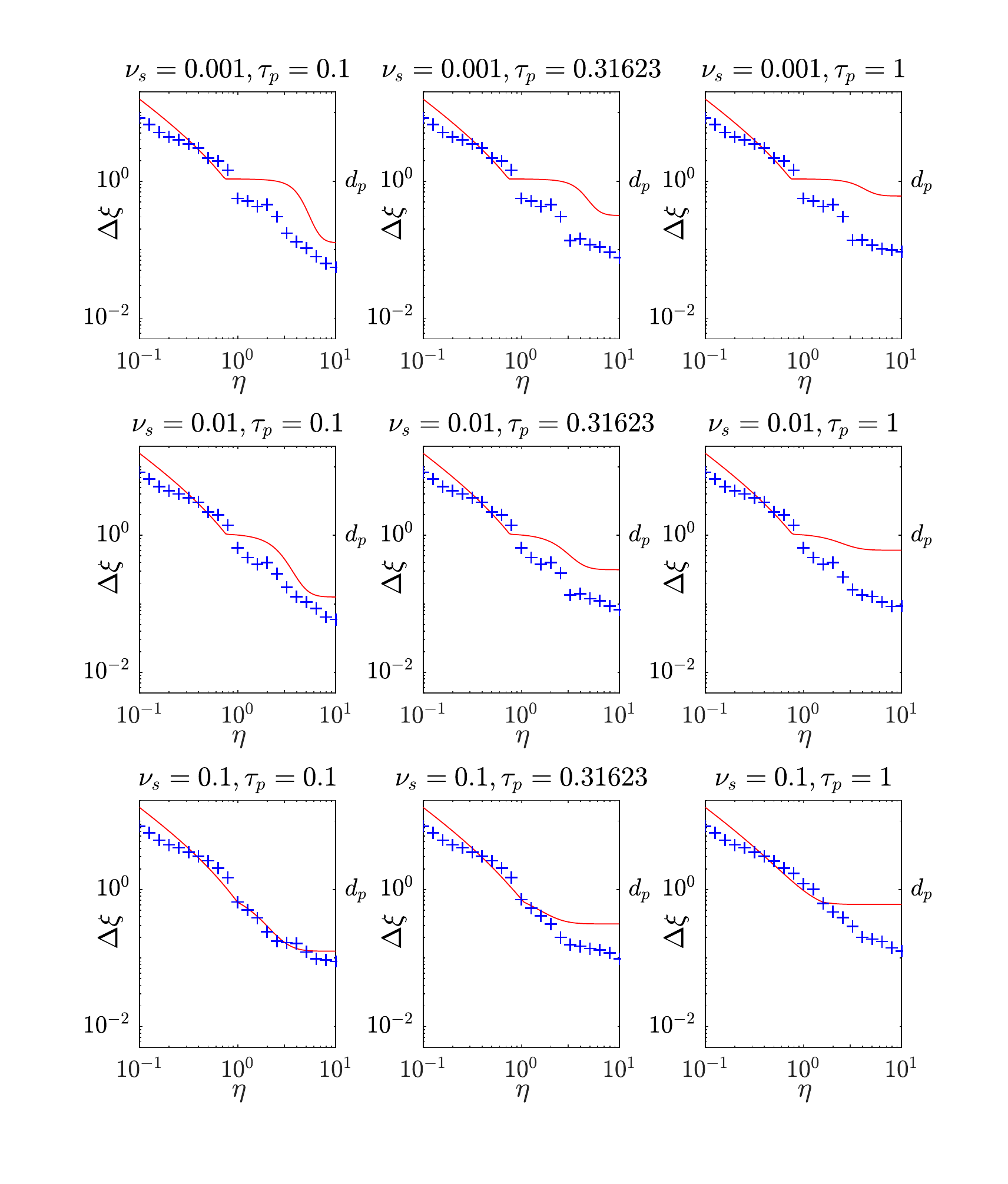}
\caption{Comparison of analytic and simulation results for the settlement front propagation width. Parameter ranges are the same as in Fig \ref{SpeedResults}. Note that in the low density limit ($\eta < 1$), the launch time $\tau_l \propto \eta^{-1}\nu_s^{-1}$ while the front speed $\nu \propto \nu_s$ so the front thickness $\Delta \xi \propto \nu \tau_l \propto \eta^{-1}$ is independent of $\nu_s$.
\label{ThicknessResults}}
\end{figure}

Figure \ref{ThicknessResults} shows the measured front thickness as well as our analytic estimate.  The analytic estimate does well in the low density regime - where the local growth following the passage of the front may well be described by a logistic curve.  However, in the high density regime the front thickness becomes less than $d_p$ and the simple exponential growth model breaks down. 

These results demonstrate that the analytic model developed in section 2 captures most of the important behavior of the settlement front in the low, intermediate and high settleable system limits.  We now use these results to estimate the crossing time of the settlement front across the galaxy.

\subsection{Galactic Crossing Time}
\begin{figure}
\includegraphics[width=1\columnwidth]{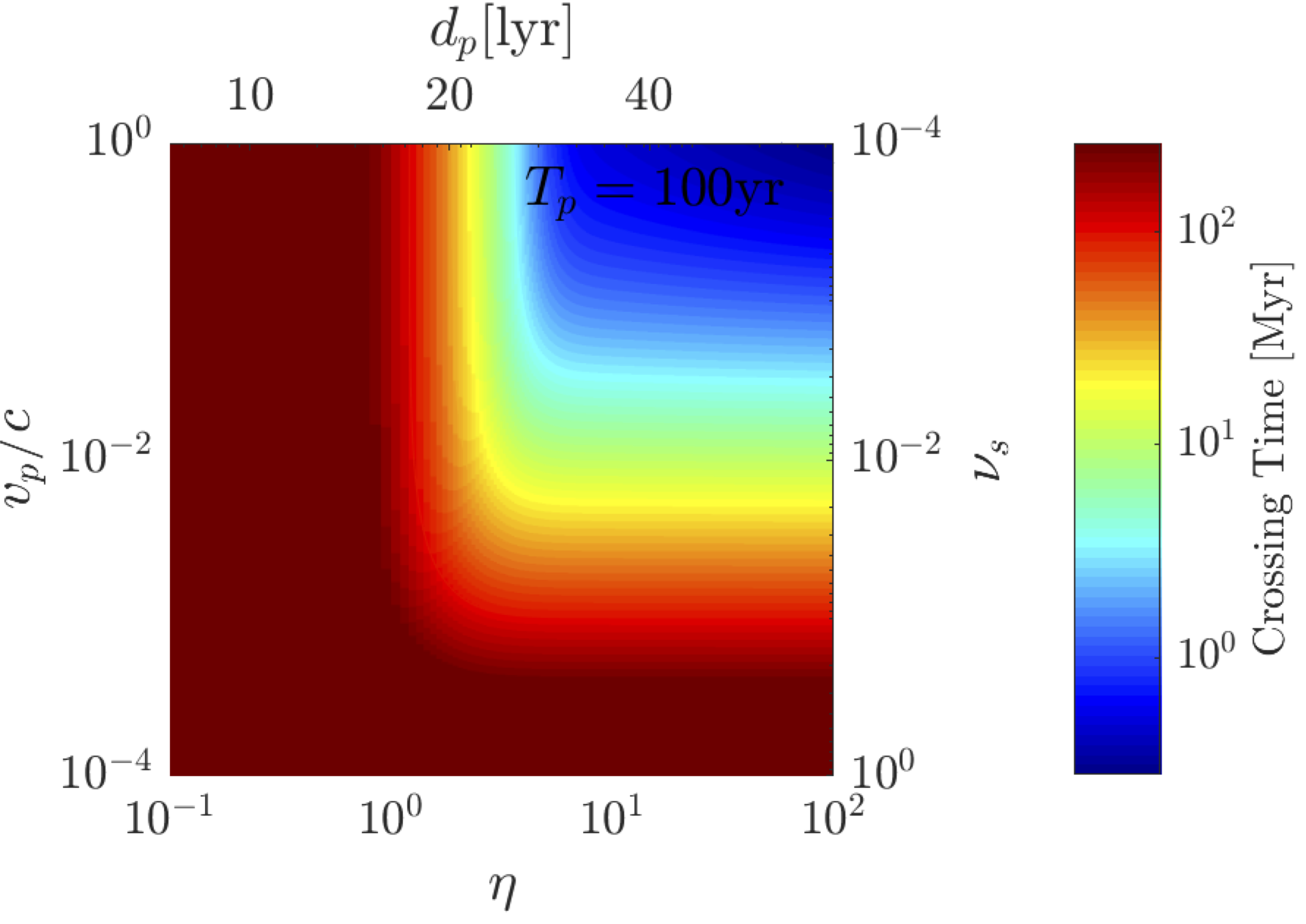}
\caption{Plot of front crossing time vs probe speed and {\settleable} system density. The plot assumes galaxy size of $10^5 \lyr$, densities similar to the solar neighborhood, stellar speeds of $30 \kms$ and a probe launch period $T_p=100 yr$.  Note that the crossing time never exceeds $300 \Myr$ which is much less than the age of the galaxy.  For reference, Voyager 1 is traveling at $\sim 10^{-4} c$. \label{galaxytime1}}
\end{figure}

\begin{figure}
\includegraphics[width=1\columnwidth]{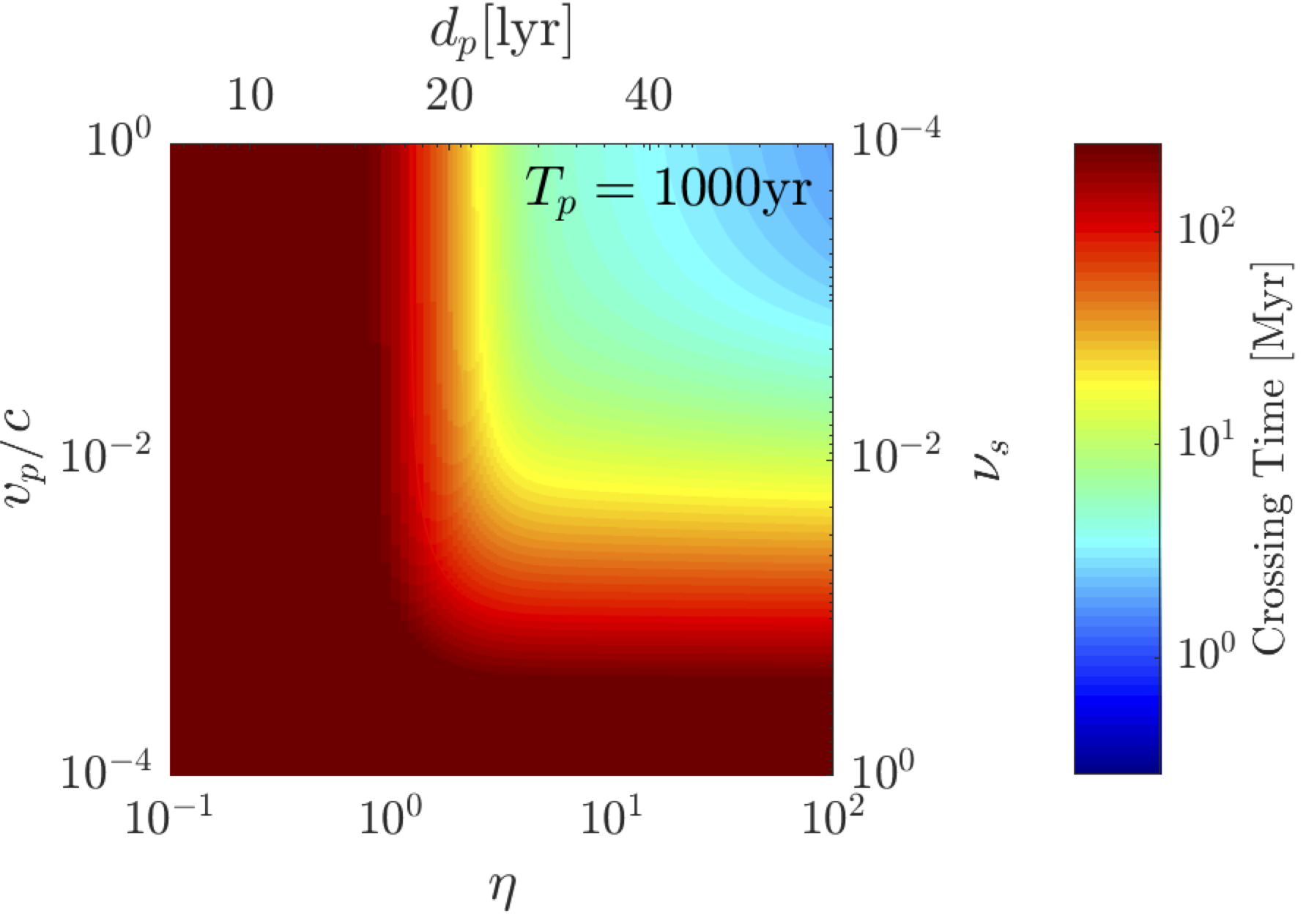}
\caption{Same as figure \ref{galaxytime1}, but with the probe launch period increased to $T_p=1000 yr$.  Increasing the time between probe launches cannot bring the settlement front crossing time above the age of the galaxy.  \label{galaxytime2}}
\end{figure}

We can now apply a very simple order-of-magnitude calculation for the Milky Way, assuming a size of $10^5 \lyr$ and a single stellar velocity dispersion of $v_s = 30 \kms$, characteristic of the Solar Neighborhood. While in reality differential rotation, motion of halo stars, and spatial variations in stellar densities and velocities will all be important corrections for realistic models of an expansion of an space-faring civilization - this speed (or rather $\mathcal{V} v_s \approx 100 \kms$)  provides a reasonable lower limit for the rate stellar motions can spread life across the galaxy interior to our galactic radius.  Using this lower limit on speed, we can calculate an upper limit on the galactic crossing time of $300 \Myr$.  This upper limit is independent of any probe speed $v_p$, {\settleable} fraction $f$, or probe range $d_p$.   Figure \ref{galaxytime1} show the galactic crossing time for a range of probe speeds and probe ranges assuming it takes $100 \yr$ to be able to relaunch a probe from a newly settled system.  Note in the low density limit $\eta < 1$, this gives $300 \Myr$.  Also as the probe speeds approach the stellar velocities $\nu_s \rightarrow 1$, the front speed becomes comparable to the stellar motions and again the galactic crossing time goes to $300 \Myr$.  If the probe speed is greater than the stellar speeds $(\nu_s < 1)$ - and the typical distances to {\settleable} systems is less than the probe range $(\eta > 1)$, the galactic crossing time approaches the light crossing time $0.1 \Myr$ as the probe velocity approaches the speed of light $(v_p/c\rightarrow 1)$.  Also note the crossing time tends to increase for shorter probe ranges in the high density $(\eta > 1)$ and high velocity $(v_p \rightarrow c)$ limit because there are more frequent hops and the time to relaunch a probe begins to become significant.  Figure \ref{galaxytime2} shows the same but for relaunch periods of $1000 \yr$.

\subsection{Fill-in Time}
While the front of settled systems may have had more than enough time to cross the galaxy, it is worth asking whether the galaxy has had sufficient time to fill in.  Starting from a single technological civilization, what is the time scale for such a civilization to grow to 100 billion civilizations?  To settle $10^{11}$ systems requires only 36 doubling times, so unless the effective probe launch time $T_l = \tau_l t_p$ is greater than $\frac{10 \Gyr}{log_2(10^{11})} = 270 \Myr$, the galaxy is old enough for every system to have been settled from an initial single civilization.  This is of order the crossing time of the entire galaxy.  To restrict launches to once every $270 \Myr$ the encounter time period $T_c$ must be greater than $270 \Myr$.  This requires $\left ( f \rho v_s \pi d_p^2 \right)^{-1} > 270 \Myr$ or using solar neighborhood densities and velocities, $ \sqrt{f} d_p <  .071\lyr$ or $4520 \AU$.  This makes close enough encounters with settle-able systems extremely rare.  

This result confirms the intuition of \cite{Brin1983}, \cite{Ashworth2012}, and \cite{Wright2014}: using realistic values for stellar motions yields galactic settlement times shorter than the age of the Milky Way, even for “slow” ships.

\section{Steady State Model}
Given that the galactic crossing time and potential fill-in time are much less than the age of the galaxy, we next consider steady state solutions for a completely settled galaxy. We assume that civilization lifetimes are finite and seek to determine under what conditions {\settleable} systems can be left unsettled for significant periods of time.  

If we assume the galaxy has had time to reach a steady state - and is homogeneous - we can model the ratio of settled to unsettled systems $(X)$ using a simple ODE.

\begin{eqnarray}
    \frac{d X}{dt} = \frac{1}{T_l} X (1-X) - \frac{1}{T_s} X\\
\end{eqnarray}
where $T_s$ is the average lifetime of settlements and $T_l$ is the effective probe launch rate. For our purposes a settlement `dies' when it ceases to be capable of launching probes.  This could be from an extinction event, resource depletion, environmental collapse or a permanent culture shift to one that does not settle nearby stars.

Note that $T_l$, the effective probe launch period, is restricted by either the time to assemble a new probe $T_p$ (in the high density limit) or by the time one would have to wait for an encounter with another {\settleable} system within the probe range.  As before we set the launch time as the weighted average of the probe assembly time $T_p$ and the collision time $T_c$
\begin{eqnarray}
    T_l =  \mathcal{D}_1 T_p + (1-\mathcal{D}_1) T_c \\
     \mathcal{D}_1=1-e^{-\frac{4 \pi }{3} \eta}
\end{eqnarray}
where $\mathcal{D}$ represents the odds of a system having at least one neighbor in range. Note we have dropped the factor of $\epsilon=\frac{1}{4}$ since in the steady state we are not concerned with advancing a front, but rather simply launching a probe to any nearby unsettled system.

The factor of $(1-X)$ represents the odds that an encounter with a {\settleable} system will be with an unsettled system.  There is a trivial steady state solution at $X=0$ which occurs if civilizations die off before they can launch any probes ($T_s < T_l$).  Otherwise, the equilibrium solution occurs at 
\begin{equation}
    X_{\eq}=1-\frac{T_l}{T_s}
\end{equation}

In equilibrium, each system must "birth" (i.e. have an encounter and settle) an average of one unsettled world in their lifetime to compensate for their own death.  If systems have several encounters with {\settleable} systems during their lifetime then on average all but one of those will be with other systems that are already settled and the average fraction of systems that are settled will be high. Note that there may be many encounters with systems which are inherently unsettleable emphasizing our use of the {\settleable} fraction $f$.  If on the other hand, systems survive just long enough to encounter another {\settleable} system (or rarely 2), nearly all of those encounters can be with a unsettled but {\settleable} system.

\begin{figure}
\includegraphics[width=\columnwidth]{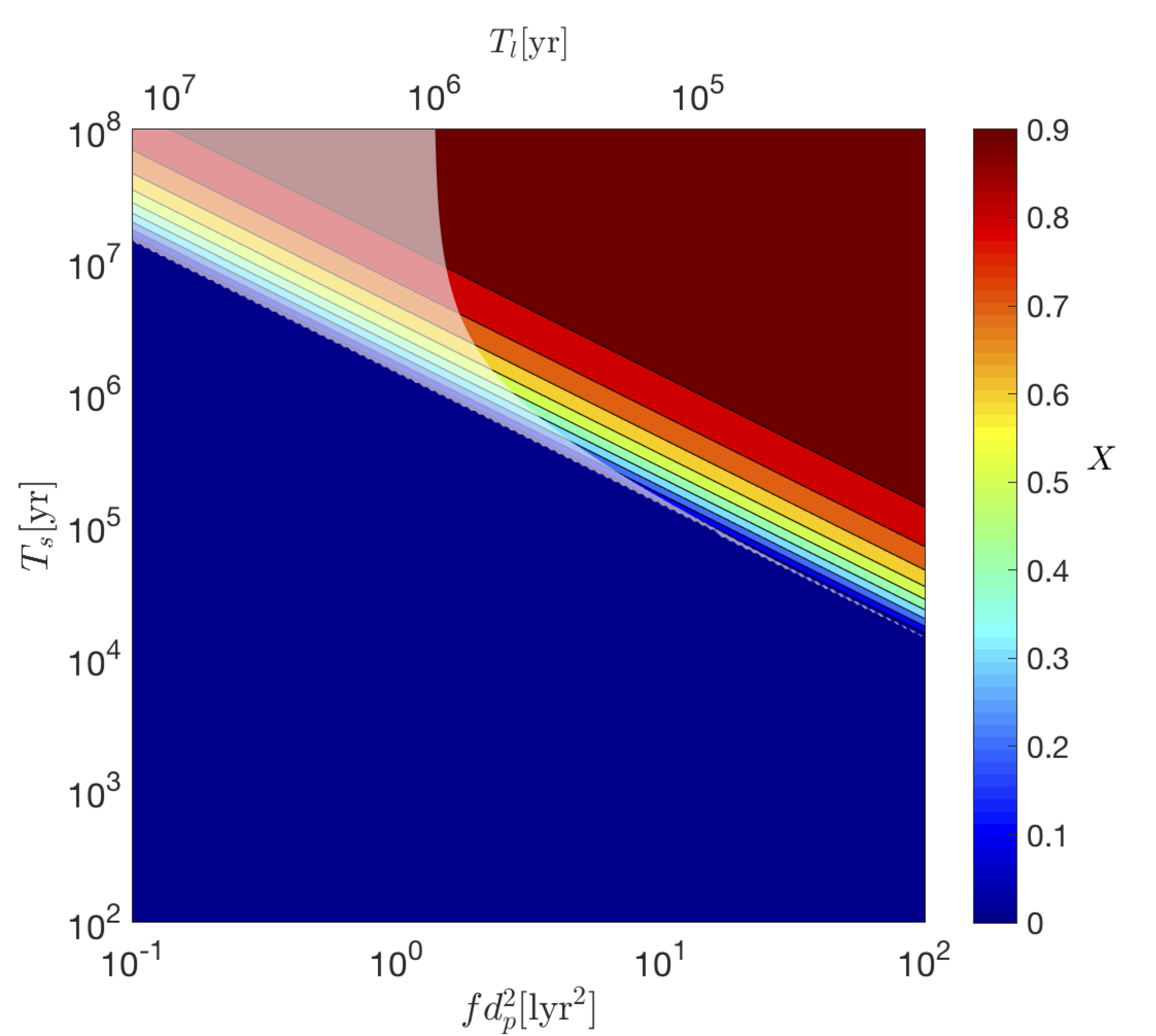}
\caption{Steady state fraction of settled systems $X$ vs settlement civilization lifetime $T_s$ and probe launch times $T_l$.  Note that in the low density limit $(\eta << 1)$, $T_l$ reduces to the time between encounters with a {\settleable} world due to stellar motions. Dark red regions are fully settled because encounters with a {\settleable} system are frequent (encounter times are less than 1 million years). Dark blue regions are sterile as encounter times are so long that civilizations die before they encounter a system to which they can send a probe. The shaded region corresponds to non-sterile conditions that could be interpreted as being consistent with Earth history (encounters times $> 1 My$: see text).
\label{LowDensityEquilibriumX}}
\end{figure}

Figure \ref{LowDensityEquilibriumX} shows the resulting equilibrium settled fraction as a function of $T_l$ and $T_s$.  In the low density limit $(\eta << \eta_c)$ the launch time $T_l$ is a function of the {\settleable} fraction $f$ and the probe range $d_p$ (assuming solar neighborhood values for density $\rho$ and stellar substrate velocities $v_s$).  This scaling is shown along the bottom axis.  In the figure dark red regions are fully settled because encounters with a settleable system are frequent. Dark blue regions are sterile as encounter times are so long that civilizations die before they encounter a system to which they can send a probe. As we will see in the following sections the shaded region of the plot corresponds to non-sterile conditions that could be interpreted as being consistent with the geologic record of Earth history (see section 1 of \cite{SchmidtFrank2018}.

\section{Steady State Simulations}

\begin{figure*}
\includegraphics[width=.5 \textwidth]{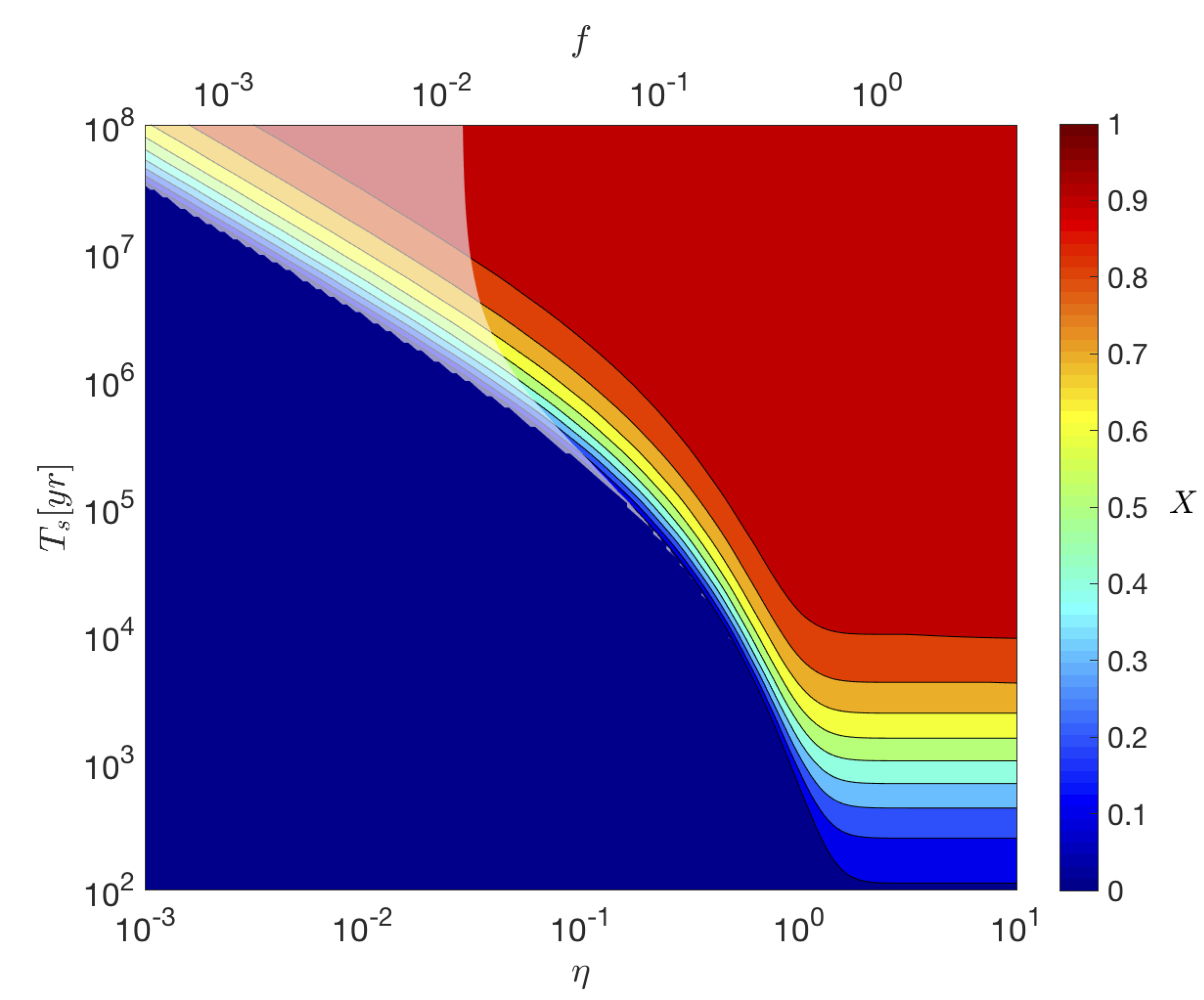}
\includegraphics[width=.5 \textwidth]{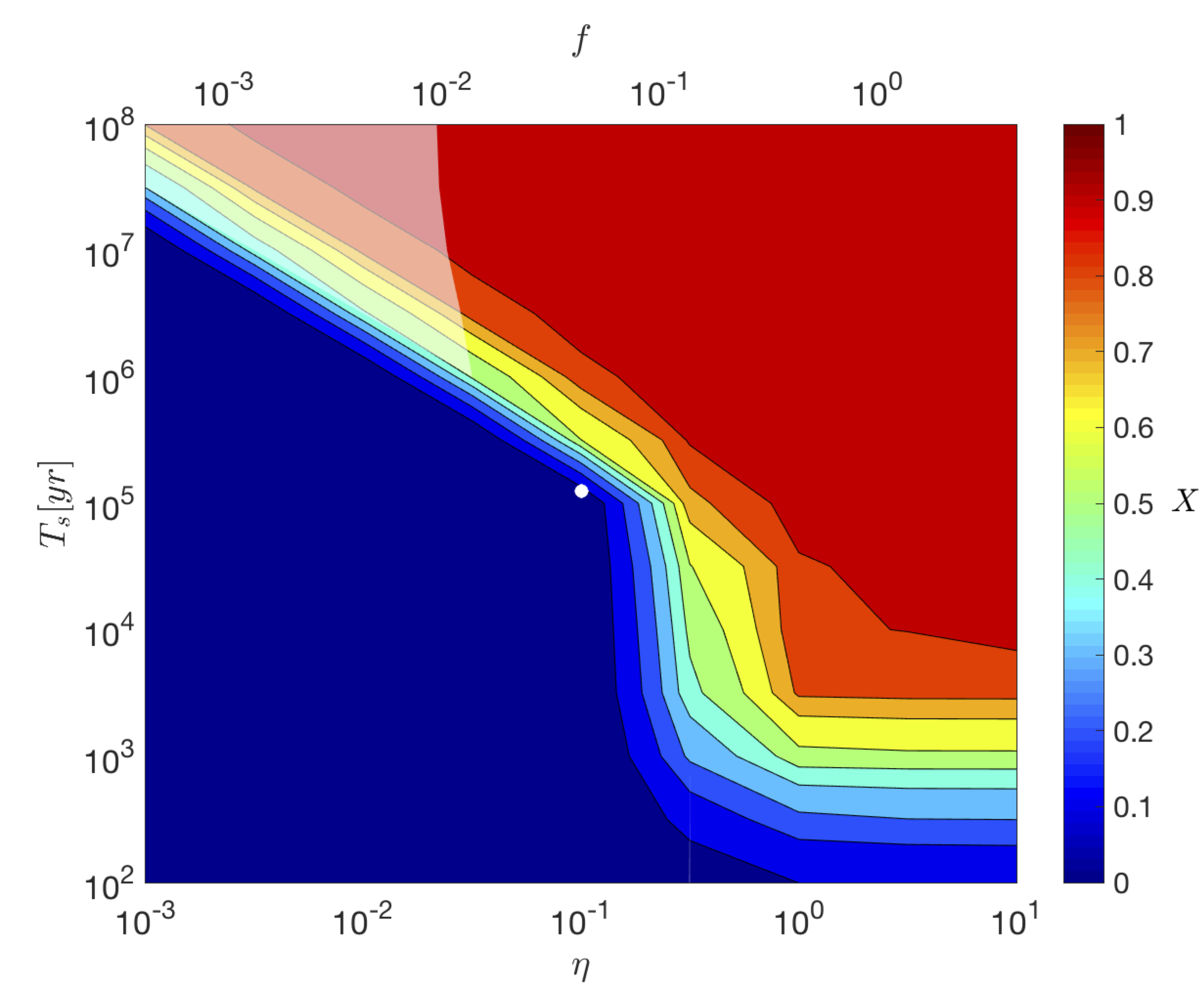}
\caption{Fraction of settled systems $X$ vs civilization lifetime $T_s$ and density of {\settleable} systems $\eta$. Red corresponds to a fully settled galaxy ($X=1$) and blue corresponds to a sterile galaxy $(X=0)$. The plot uses solar neighborhood densities and stellar velocities, a probe range of $10$ \lyr and velocity $.01 c$. $f$ is the fraction of {\settleable} systems. The left panel uses the analytic solution while the right is calculated from an array of simulations \label{EquilibriumFractionResults}. The white dot corresponds to the snapshot shown in figure \ref{ClusteringSnapshot}.  Note the transition in the settled fraction as a function of density (around $\eta \sim 1$ and $f \sim 1$). For $\eta > 1$ full settlement $X=1$ might be expected, however short lifetimes can produce low settlement fractions.  Note also that the simulations show a higher than expected settlement fraction for densities $\lesssim 1$. This is due to local statistical fluctuations producing over abundances of {\settleable} worlds.  These continue to resettle each other after the parent civilization dies. }
\end{figure*}

To validate the model we ran a sequence of 117 simulations using parameters for the solar neighborhood and explored the dependence on the final settled fraction $X$ on the fraction of {\settleable} systems $f$ and the lifetime for civilizations $T_s$.  We took a constant probe range of $10 \lyr$ and a probe speed of $.01 c$ and set the probe launch period $T_p = .01 \min(T_s, T_c)$.  Each simulation contained  $N=10^4$ particles in a periodic box with a Maxwellian distribution of velocities.  We ran each simulation for $20 T_c$, which was long enough for the settled fraction $X$ to have reach a steady state.  Initially the fraction of settled systems was set to $\max\left [ .01, X_{\eq} \right]$ randomly distributed throughout the box. This was sparse enough that at early times these initial systems evolve essentially independently. This means one would need need ever larger samples to study the behavior for lower values of the initial settled system population ($X_{\eq} << 1$).

We then looked at the equilibrium settled fraction (averaged over the final $T_c$) compared with our analytic model.  The right panel of figure \ref{EquilibriumFractionResults} shows the resulting settled fraction.  In the high density limit $(\eta > 1)$, $T_l = T_p = .01 T_s$ and we would naively expect $X_{\eq}=.99$, however the measured settled fractions are lower.  This is because in the numerical models, probes are not launched towards systems until after they are no longer settled.  And if the travel time is longer than the lifetimes, this delay can create many more targeted systems than settled systems.  The appendix contains a modified model that accounts for this delay and its predictions are shown in the left panel of figure \ref{EquilibriumFractionResults}.  At both low and high densities it agrees fairly well with the numerical results, though the transition to higher settled fractions appears to be much broader and happen much sooner around $\eta \approx .2$ as opposed to $\eta_c$.  This is likely due to a degree of back-filling discussed below.  

\subsection{Resettlement in the Numerical Models}

The numerical models showed two interesting phenomena.  First, many of the models with $T_s << T_l$ continued to have a handful of systems ($N<6$) survive for very long times.  This was due to pairs of systems very close in phase space with similar velocities and positions that were able to resettle each other over much longer timescales than their own lifetimes.   Looking at the number of unique systems settled over the latter half of the steady state simulations, we found a clear break between those with only a handful of systems  $(N=6$ or $.06\%)$, and those that used a substantial fraction of the systems $(48\%)$.  While this back and forth allows for settled systems to exist in the numerical model, we consider any simulation that survives by reusing fewer than 6 systems, as having a settled fraction of 0. 

The other interesting phenomena was the excess of systems   visible in the right panel of figure \ref{EquilibriumFractionResults} below $T_s = 10^5 \yr$ for $\eta$ between $.1$ and $1.0$. This too is likely due to an increased probability of resettling reducing the effective launch period in regions with above average density approaching the critical density.  For $\eta = \eta_1 = .25$, systems typically have a single neighbor in range allowing for pairs of systems to resettle one another even if their lifetimes are very short.  By $\eta = \eta_2=.4$ there can be subgroups containing $\approx 10$ systems that can continually resettle one another.  Within these groups, the probe launch time is just the probe assembly time and if a system dies it can be resettled by nearby systems.  These pockets of settled systems can settle other systems that migrate through. These become the seeds for other pockets to arise if encountered before the system's lifetime.  And by $\eta = \eta_4 = .88$ systems are fully connected.

\begin{figure}
\includegraphics[width=\columnwidth]{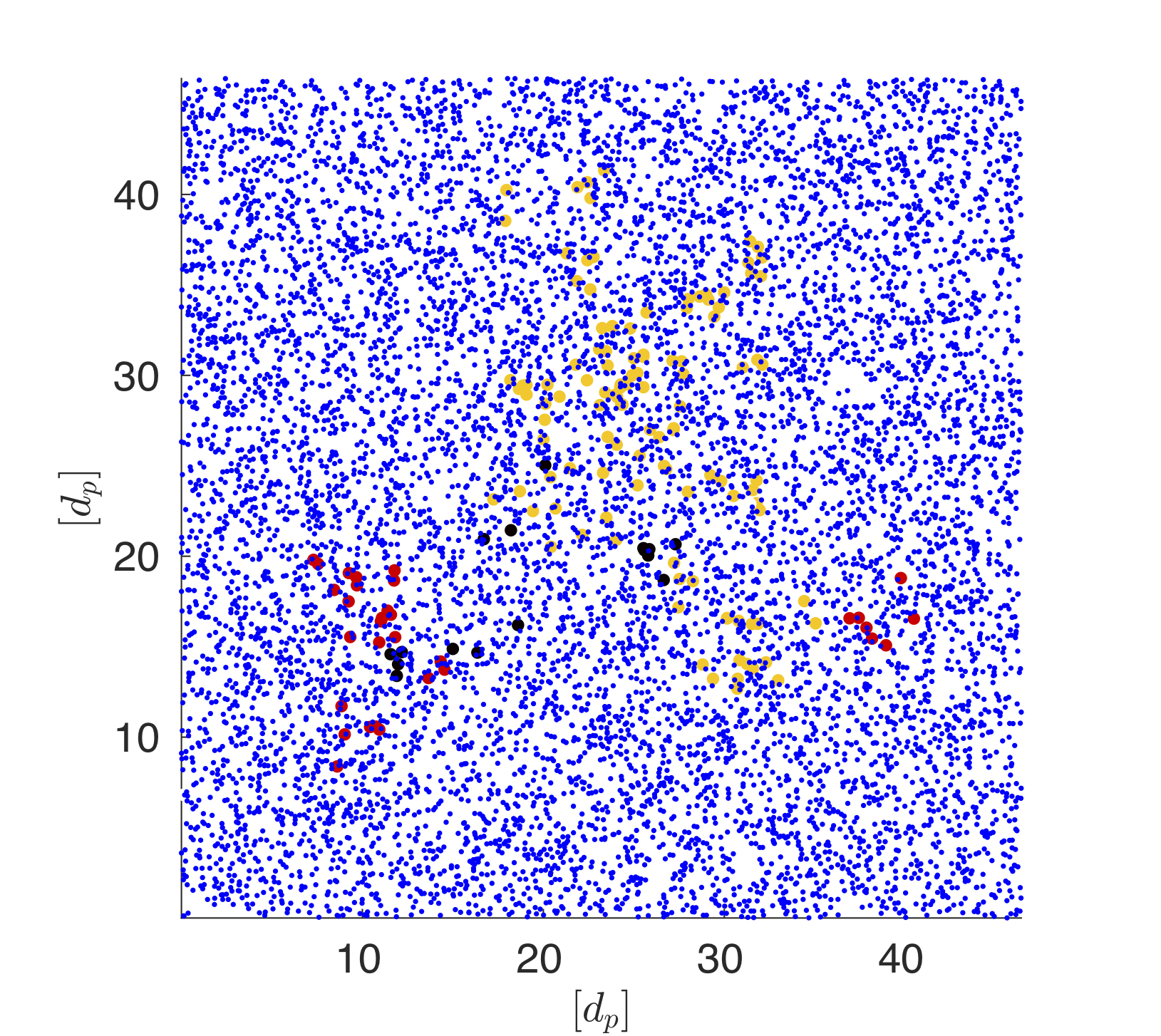}
\caption{Settlement simulation snapshot. Blue dots are unsettled systems. The colored circles show settled systems. Systems with the same color share a common ancestor.  For these conditions we expect $X \sim 0$ because the effective launch time is longer than the civilization lifetime ($T_l/T_s \sim 2$). Clusters persist, however, because local statistical fluctuations produce over abundances of {\settleable} worlds which can continually resettle one another.
\label{ClusteringSnapshot}}
\end{figure}

Figure \ref{ClusteringSnapshot} shows the results from a run with a settlement lifetime $T_s = 1.25 \times 10^5 \yr$ and a encounter time $T_c = 3.18 \times 10^5 \yr$.  The density $\eta = .1$ gives a neighbor probability $\mathcal{D}_1=.342$ and an average launch time $T_l = 2.1 \times 10^5 \yr$.  While this is still longer than the settlement lifetime, implying $X_{\eq}=0$, a significant number of systems are able to survive in local pockets with higher than average settlement fractions.  This is because settled systems (having just been settled) are more likely to have neighbors in range - some of which will have recently become unsettled.

\begin{figure*}
\includegraphics[width=.5 \textwidth]{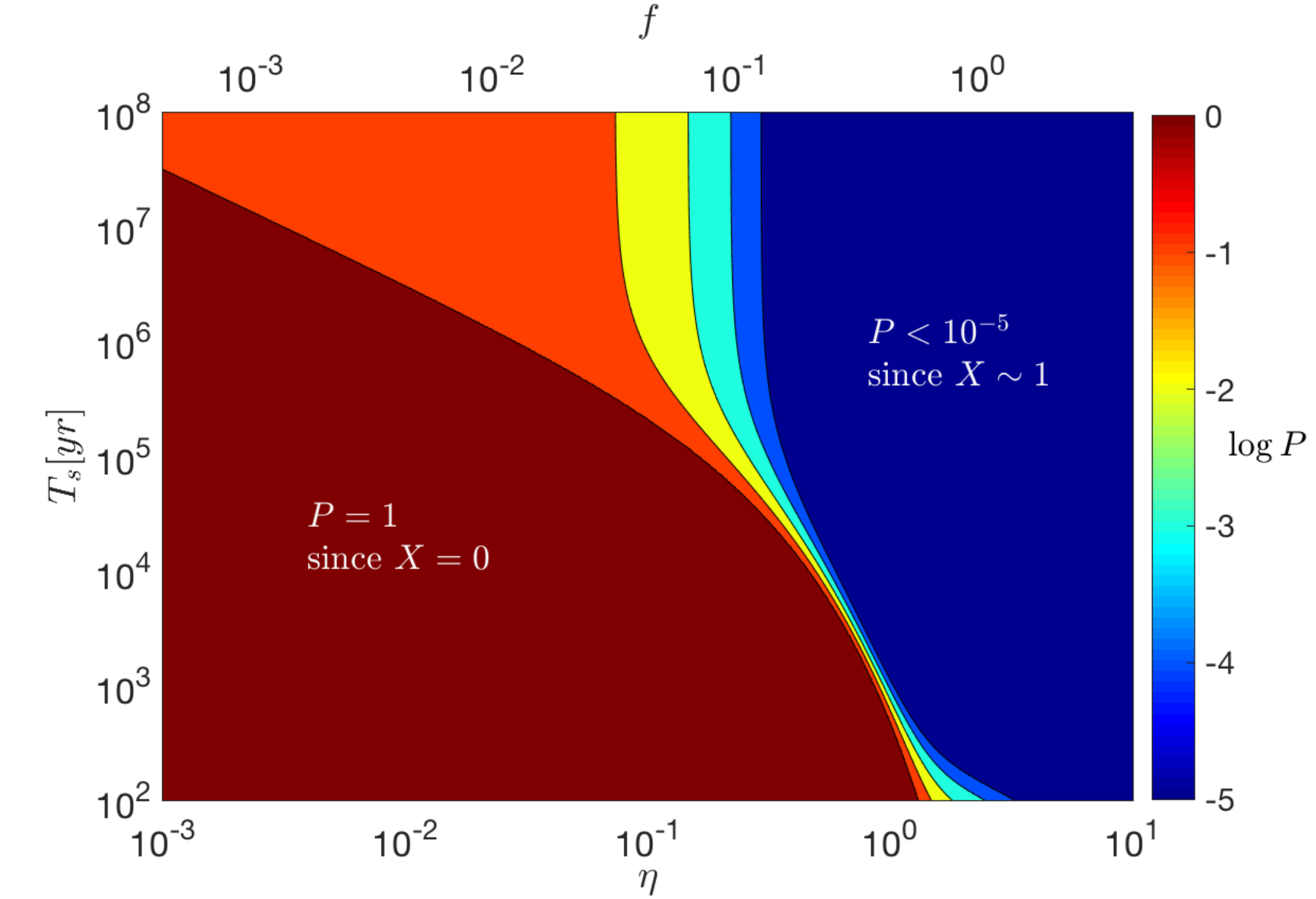}
\includegraphics[width=.5 \textwidth]{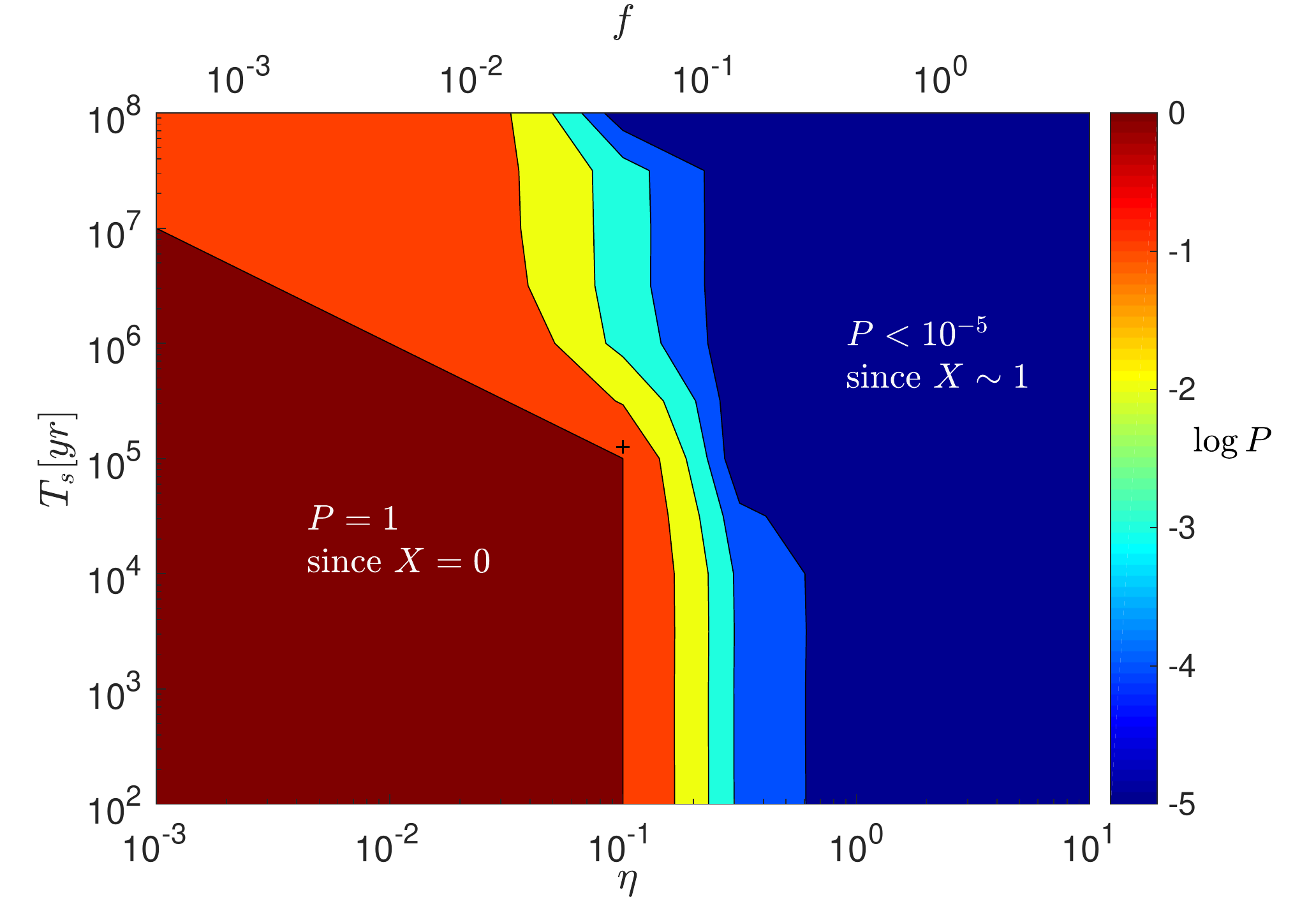}
\caption{Equilibrium Galactic Settlement vs Earth Observation Constraints. Here we show the probability of systems being unsettled for at least $1 \Myr$ vs civilization lifetime $T_s$ and the density of {\settleable} systems $\eta$.  Left panel shows analytic model results. Right panel shows simulation results.  The dark red regions have a probability of 1 because $X=0$, meaning there are no space-faring civilizations which explains why Earth has not been settled in the last $1 \Myr$. The absence of civilizations occurs because they die out before being able to encounter another {\settleable} system.  The dark blue regions have a very small probability of going $1 \Myr$ without encountering another space faring civilization. This is because most systems are settled and densities are high enough that encounter times are short.}
\label{NoAliens1e6}
\end{figure*}

\section{Evidence Horizons and Hart's Fact A}
We know turn our attention to Hart's Fact A which focuses on the question of why Earth has not, apparently, been settled (or at least visited) by another space faring civilization.  The important point to consider is the temporal one. How long ago could Earth have been (temporarily) visited or settled by such a civilization without leaving any obvious trace? If the settlement occurred 4 billion years ago and lasted for just 10,000 years would any record of it survive in the geological record? 

The answer is: almost certainly not. This implies a temporal horizon over which a settlement might not be ``seen". We now use the equilibrium solutions to our steady state model and attempt to constrain the reasoning used in linking Fact A to conclusions about Fermi's Paradox.  

Given assumptions about probe range $(10 \lyr)$ and velocity $(.01 c)$, we now calculate the typical launch time and equilibrium fraction as a function of the fraction of {\settleable} systems $f$ and the average settlement lifetime $T_s$ (assuming solar neighborhood density of systems and stellar velocities).  If encounters between unsettled and settled systems are uncorrelated, we would expect the distribution of times a system remains unsettled $(T_u)$ to follow a Poisson distribution

\begin{equation}
    P(T_u)=e^{-\frac{T_u}{T_l}X}
\end{equation}

We can then inquire about the probability of being unsettled for some period which can be compared against evidence in Earth's geological record. \cite{SchmidtFrank2018} discuss the mechanisms by which evidence of a previous technological civilizations on Earth  would be problematic to find and would likely only exist as potentially ambiguous chemical or isotopic signals, if at all. Motivated by the maximum time span of our simulations, we choose this horizon to be at least $1 \Myr$. Note that this is a lower limit in that our results are relevant to being unsettled for at least this long. As the horizon time gets longer, the lighter red region in Figure \ref{EquilibriumFractionResults} that allows for a reasonable probability of $X >0$ moves upwards toward longer settled times $T_s$ and to the left towards lower fractions of {\settleable} systems $f$.

The resulting probability of a given system being unsettled for at least $1 \Myr$ is plotted in figure \ref{NoAliens1e6} vs.\ civilization lifetime $T_s$ and the density of {\settleable} systems $\eta$.  The left panel shows results for the analytic model; the right panel shows results from the simulations, which validate the analytic model over most of the parameter space we have explored.

The dark red regions of Figure \ref{NoAliens1e6} show probability of 1 that the Earth has not been recently visited because $X=0$ (i.e.\ because there are no space-faring civilizations, because  civilization lifetimes are shorter than the time to encounter another {\settleable} system.)  The dark blue regions (indicating a very small probability of Earth having gone $1 \Myr$ or more without encountering another space-faring civilization represents parameters pace where most systems are already settled,  and the densities of {\settleable} systems are high enough that encounter times between them and the Earth are short.

The most important point we draw from these plots is that between these two possibilities lies a range of conditions in which the galaxy supports a population of interstellar civilizations even though Earth would likely not experience a settlement attempts for $1 \Myr$ or more. Consider, for example, the situation with a settlement fraction $f = 0.03$ and civilization lifetime of $10^6$.  In this case we find it is not terribly unlikely ($P \sim .1$) that Earth has remained unvisited for the past $1 \Myr$ or longer, even though $X > 0$.  As we will discuss, this result has important conclusions for interpretations of Hart's Fact A and the Fermi Paradox. Also we note that as the look-back time is extended the analytic version of plots like those shown in figure \ref{NoAliens1e6} tend to have smaller intermediate probabilities for short-lived settlements. We expect however that longer run simulations would retain their wider range in probabilities. We note that the question of ``evidence horizons" should be the focus of its own in-depth future study.

Finally, it is worth pointing out a few features of the probabilities found in our steady state numerical models (right panel of figure \ref{NoAliens1e6}).  We found a small probability of Earth remaining unsettled for the past $1 \Myr$ or longer for $T_s < 10^5 \yr$ for values of $\eta \sim .3$ primarily due to small equilibrium values of $X$ supported by resettling near the critical density.  This occurs without balancing $T_s$ and $T_l$. The black "+" in the right panel of figure \ref{NoAliens1e6} represents a run in which a few clusters containing $10$'s of systems tended to persist.  In this run the odds of going $1 \Myr$  without a settlement was $\sim 89\%$. Thus if the parameters in this run where to represent the situation in our region of the galaxy and Earth was not in one of the ``re-settlement" clusters it would be highly probable that we would not have been settled (or visited) by another civilization for at least $1 \Myr$.  
 Note also that the narrow transition from $P=10^{-4}$ to $P=1$ for $T_s < 10^5 \yr$ between $\eta=.1$ and $\eta=.3$ seen in the left panel of figure \ref{NoAliens1e6} could not be resolved by our grid of runs. The width of the transition seen in the contour plot on the right panel roughly corresponds to the resolution in $\eta$ of our grid of simulations and could not be any narrower given the interpolation used to generate the contours.

\section{Discussion and Conclusions}

\subsection{Conclusions}

We now review and discuss the principle conclusions of our work. We can summarize our conclusions as follows
\begin{itemize}
    \item When diffusive stellar motions are accounted for, they contribute to the Galaxy becoming fully settled in a time less than, or at very least comparable to its present age, even for slow or infrequent interstellar probes.
    \item If a settlement front forms, all {\settleable} systems behind it become ``filled in" in a time less than the current age of the Galaxy.
    \item While settlement wave crossing and fill-in times are short, consideration of finite civilization lifetimes in a steady state model allows for conditions in which the settled fraction $X$ is less than 1. Thus the galaxy may be in a steady state in which not every {\settleable} system is currently settled.
    \item Even for regions of parameter space in which one might expect $X \sim 0$ for typical regions of the Galaxy, statistical fluctuations in local density of {\settleable} systems allows for the formation of settlement clusters which can continually resettle one another.  These clusters are then surrounded by large unsettled regions. If such conditions represent the situation in our region of the galaxy and Earth was not in one of the "re-settlement" clusters it would be highly probable that we would not have been settled (or visited) by another civilization for some time.
    \item By consideration of the convolution of steady state solutions with geologic evidence horizons, it is possible to find situations in which Earth may not have experienced a settlement event for longer than some horizon time (set to 1 Myr in this work) even though the galaxy supports a population of interstellar civilizations.
\end{itemize}

Our first conclusion shows that if diffusive stellar motions are accounted for it appears almost unavoidable that if any interstellar-space-faring civilization arises, the galaxy will become fully settled in a time less than, or at very least comparable to its present age.  In particular, we confirm that thermal motions of stars prevent settlement fronts from ``stalling'' for timescales longer than the age of the Milky Way, as suggested by \cite{Brin1983, Ashworth2012, Wright2014}.  Thus if the practical and technological impediments to interstellar settlement are overcome, then the ``wave" of settlement should sweep across the entire Galaxy.

Note that we find that the settlement front crossing time and fill in time takes of order $\lesssim 1 \Gyr$ even for “slow” probes $(30 \kms)$. This speed is significant because it corresponds to typical interplanetary probe speeds we can design today, and is of order the speeds a ship of any size can achieve via gravitational slingshots with giant planets in ~1 AU orbits.  

Our conclusions on the settlement time of the Galaxy are almost certainly lower limits for two reasons. The first is that we have not included the effects of galactic shear or halo stars in our simulations, and these will provide additional opportunities for ``mixing” in the case of slow ships that will cause the settlement front to expand faster than the speeds of the ships themselves.

The second is that we have assumed zero variation or improvement with time in probe launch rates, probe ranges, probe speeds or exo-civilization lifetimes. A more realistic description of spaceflight technology on Gyr timescales would include variation among the settlements, and the expansion would likely be dominated by the high-expansion-rate tail of this distribution.

Our third conclusion concerns steady states for galactic settlement.  Allowing for civilizations to have finite lifetimes, the steady state fraction of settled worlds will be a function of both civilization lifetime and the rate of encounters with ``empty" unsettled worlds. Our results show that there are regions of the parameter space where $0<X<1$. Thus our steady states model quantifies a possibility not generally considered in previous discussions of the Fermi Paradox and Galactic settlement.  It is possible to achieve a galaxy in which there remain unsettled worlds even though the settlement front has crossed the entire galactic disk.  

Our fourth conclusion comes from numerical simulations of the steady state model.  Here we find under some conditions the encounter times can be so long that we would expect $X = 0$ throughout the galaxy.  This would occur because civilizations would always die out before a settlement opportunity occurred.  Our simulations show however that local statistical over-abundances of {\settleable} systems can occur.  This leads to clusters of closely-packed settled systems surrounded by larger unsettled voids.  If Earth were to exist within one of these voids then it mean that there was a high probability that Earth might never have experienced a settlement event.

Our final conclusion concerns the temporal aspect of the Fermi Paradox and Hart's Fact A---the lack of any obvious settlement of the Solar System---which Hart argues compels the conclusion that there can be no other technological civilizations in the galaxy.  By including a finite time horizon past which evidence of prior settlement civilization might not be seen, we have shown that it is possible to break the link between Hart's Fact A and his conclusion.  To wit, it is possible to have a galaxy with some non-zero settlement fraction and still have evidence for a prior Earth ``visit" lie over the horizon available via Earth's geological record.  Indeed, the last three of our conclusions all break the link between conclusions about rapid galactic settlement and the current absence of technological civilizations.  This occurs because the steady-state model implies that not all {\settleable} systems need be currently occupied. 

\subsection{Discussion}

For low densities of {\settleable} systems, our steady-state calculation finds that consistency with the lack of evidence for Earth's past settlement requires that each civilization has, on average, only one chance to reproduce i.e. to settle another world.  This does not mean that only one settlement probe was launched however.  Our result can be interpreted to mean that on average only one settlement probe was successful. Inherent in our calculation was the settlement fraction $f$. Our steady state calculation was carried out in the low density limit which implies $f < 1$ (note that in the high density limit the fraction simply tends to $X=1$).  As described earlier, low values of $f$ implies ``good planets are hard find".  Our steady state calculations in the low density limit further imply that ``successful settlements are hard to achieve". The lack of settlement success could come for many reasons ranging from failure of interstellar vessels capable of establishing persistent settlements to the inability to develop viable progeny civilizations on new worlds.

Because Hart's conclusions stem from his assumption that we would have noticed if extraterrestrial technology had ever settled the Solar System, they are challenged by the work of \citet{Freitas1985}, \cite{SchmidtFrank2018}, \cite{Davies2012}, and \cite{Haqq-Misra2012} who show that this is not necessarily the case. 

We can go further though: Hart's conclusions are also subject to the assumption that the Solar System would be considered {\settleable} by any of the exo-civilizations it has come within range of. The most extravagant contradiction of this assumption is the Zoo Hypothesis \citep{Ball1973}, but we need not invoke such ``solipsist'' positions \citep{Sagan1983} to point out the flaw in Hart's reasoning here. One can imagine many reasons why the Solar System might not be {\settleable} (i.e.\ not part of the fraction $f$ in our analysis), including the {\it Aurora} effect mentioned in Section \ref{Introduction} or the possibility that they avoid settling the environment near the Earth exactly because it is inhabited with life.

In particular, the assumption that the Earth's life-sustaining resources make it a particularly {\it good} target for extraterrestrial settlement projects could be a naive projection onto exo-civilizations of a particular set of human attitudes that conflate expansion and exploration with conquest of (or at least indifference towards) native populations \citep{WrightOmanReagan2018}. One might just as plausibly posit that any extremely long-lived civilization would appreciate the importance of leaving native life and its near-space environment undisturbed.

So our conclusions have strong implications for the likelihood of success of SETI, but the specific nature of that optimism is strongly dependent on assumptions regarding either the limits of technology or the agency of exo-civilizations. If large \-scale terraforming is not a realistic possibility then $f$ may be limited by worlds matching the biology of the parent or Ur civilization. Earth may simply not be one of those worlds even though such settlements exist elsewhere.  Likewise if, for whatever reason, all extraterrestrial civilizations that have had to the opportunity to settle the Solar System have avoided it, then the Milky Way should be filled with stars hosting potentially detectable extraterrestrial technology if even a single settlement front has ever been established. 

If instead one follows Hart and assumes that Solar System settlements would be inevitable, then our analysis quantifies the regions of parameters space consistent with his Fact A in which the Galaxy is filled with, or devoid of, space-faring technology. Those sets of parameters in which Fact A does imply an ``empty'' Milky Way are those in which Galactic settlement is especially efficient.  This implies that {\it other} galaxies where technological life has arisen should have been thoroughly settled, raising the prospects that they might be detected at extragalactic distances.

We note also that in the classic argument, Hart's Fact A is linked to conclusions about exo-civilizations because it is assumed that interstellar travel is a natural result of their evolution. But this need not be the case.  In \cite{Ashworth2012}, the energetics of developing interstellar ships that could host long term viable populations was explored.  Given the travel times between stars, these would be multi-generational `world ships' \cite{Ashworth2012} attempted to calculate the cost of building such machines, including factors such as speed and mass.  His finding was that that economies equivalent to that of entire solar systems would be required to develop and launch world ships.  As an example, consider his `medium multi-generational cruiser' case.  This was ship traveling at $v = 0.05c$, carrying a population of $10^4$ people and weighing $10^7$ tonnes.  Such a ship would require a power of 6900 zettajoules (ZJ).  He estimates that a solar system wide civilization of 900 billion people would generate 1136 ZJ per year.  Thus while the creation of world ship by such an economy would be possible, it would require a significant proportion the civilization's resources.  We note that these estimates are, of course, highly speculative and \cite{Ashworth2012} also provides estimates for solar system wide civilizations generating even higher power economies.  

For our present results these factors indicate that it is possible that developing the requirements for interstellar settlement may be expensive enough to be universally prohibitive.  In addition, if establishment of viable settlements proves difficult, meaning the success rate of world ships is low, then civilizations may be unwilling to continue investing in them over time.  This is particularly true if one considers that the long travel and communication times may make it difficult to establish an interstellar civilization.  Unless the individuals in the species driving the settlement have very long lifetimes ($> 100$ y) it is difficult to see how a galactic scale culture can arise (i.e. commerce etc. \cite{Krugman2010}). Thus each settlement may, in practice, be relatively isolated culturally which may limit the effort civilizations are willing to put into long term programs of expansion.

It is also worth considering the distribution of natural catastrophes which might lead to end of settlements.  Weak constraints might be obtained using Earth as an example by looking for cross-correlation with the ages of the impact craters, super-volcanic deposits and extinctions. The timescales between such events is likely longer than 1 My and more work can be done to explore the question of look-back horizons for evidence of settlement events.  As noted earlier, as the look-back time is extended plots like those shown in Figure \ref{NoAliens1e6} tend to have smaller intermediate probabilities for short-lived settlements.

In summary: our work demonstrates that even though settlement fronts can be expected to cross the galaxy quickly, every {\settleable} system need not be inhabited.  We note that much work needs to be done to extract the maximum amount of information from this fact when convolved with both the expected conditions for different regions of the galaxy along with what can plausibly be expected from Earth's geologic record.  In particular further studies of the settlement steady states may help understand the creation, extent and longevity of settlement voids which provide one explanation for the lack of evidence for Earth's past settlement. Our calculations open a new avenue in consideration of exo-civilizations and their prevalence in the Galaxy, and have strong but assumption-dependent implications for the prospects of the success of SETI.

\section{\bf Acknowledgements}
We thank Milan M.\ \'Cirkovi\'c, Woody Sullivan, Gregory Van Maanen, Andrew Siemion and David Brin for their helpful discussions.  We also acknowledge the New York Academy of Sciences for hosting a panel discussion which helped create the impetus for this work. CS acknowledges support from the NASA Astrobiology Program through participation in the Nexus for Exoplanet System Science and NASA Grant NNX15AK95G
The Center for Exoplanets and Habitable Worlds is supported by the
 Pennsylvania State University, the Eberly College of Science, and the
 Pennsylvania Space Grant Consortium.  The authors thank the Center for Integrated Research Computing (CIRC) at the University of Rochester for providing computational resources and technical support.


\section{Appendix}
\subsection{Calculating $\mathcal{V}(n)$}
The expectation value for the maximum velocity from $N$ samples from a Maxwell-Boltzmann distribution
\begin{equation}
    \mathcal{V}(n) = E[V_{\max}(n)]
\end{equation}
where $V_{\max}$ is the distribution of maximum values from a sample of $n$ velocities taken from a velocity distribution $P(v)$.  For a given velocity distribution $P(v)$ with cdf $C(v)$, the odds that n random samples are below a particular value $v_0$ is  $C(v_0)^n$. This is then the cumulative distribution function for the maximum value $C_{\max}(v)$.  To get the expectation value, we have $E[v]=\int v \times P_{\max}(v) dv$ where $P_{\max} = \frac{d C_{\max}(v)}{dv} = \frac{d C(v)^n}{dv}$.  So we have

\begin{equation}
    \mathcal{V}(n)=\int v \frac{ d (C(v)^n)}{dv} dv
\end{equation}

\subsection{Including targeted systems}

If we assume that settled systems do not launch probes at other systems that are already settled, we can modify our model by including the fraction of systems that are targeted in addition to those that are settled and habitable.  

\begin{eqnarray}
    \frac{d N_s}{d t} = \frac{1}{t_p} N_t - \frac{1}{T_s} N_s\\
    \frac{d N_t}{d t} = \frac{1}{T_l} N_s N_h - \frac{1}{t_p} N_t \\
    \frac{d N_h}{d t} = \frac{1}{T_s} N_s - \frac{1}{T_l} N_s N_h \\
\end{eqnarray}

where $N_s$ is the number of settled systems, $N_t$ is the number of targeted systems, and $N_h$ is the number of habitable systems.  In addition $t_p$ is the probe travel time, $T_s$ is the settled time, and $T_l$ is the effective launch time (due to either probe production rates or encounter times).  We can dimensionalize the equation where $X = \frac{N_s}{N}$, $Y = \frac{N_t}{N}$ and $\frac{N_h}{N} = 1-X-Y$.  The resulting equations are 

\begin{eqnarray}
    \frac{d X}{d t} = \frac{Y}{t_p} - \frac{1}{T_s} X\\
    \frac{d Y}{d t} = \frac{1}{T_l} X (1-X-Y) - \frac{Y}{t_p} \\
\end{eqnarray}

Solving for equilibrium, we find
$X=\frac{T_s-T_l}{T_s+t_p}$


\end{document}